\documentclass[twocolumn,showpacs,preprintnumbers,amsmath,amssymb,superscriptaddress,longbibliography]{revtex4-1}   

\usepackage{graphicx}
\usepackage{dcolumn}
\usepackage{empheq}
\usepackage{amsmath,mathrsfs,amssymb,amsthm}  
\usepackage{hhline} 
\usepackage{wasysym,enumerate,color}
\usepackage{epstopdf} 
\usepackage{verbatim}  
\usepackage{bm}
\usepackage{ulem} 
\usepackage{color}
\usepackage[colorlinks=true,citecolor=blue,linkcolor=blue,urlcolor=blue]{hyperref}
\definecolor{darkgreen}{rgb}{0,0.5,0}
\definecolor{orange}{rgb}{1,0.5,0}
\definecolor{grey}{rgb}{.6,.6,.6}

\newcommand{\e}{\varepsilon}
\newcommand{\s}{\sigma}

\newcommand{\be}{\begin{equation}}
\newcommand{\ee}{\end{equation}}
\newcommand{\beq}{\begin{eqnarray}}
\newcommand{\eeq}{\end{eqnarray}}

\newcommand{\fig}[1]{Fig.~\ref{#1}}
\newcommand{\eq}[1]{Eq.~(\ref{#1})}
\newcommand{\exch}{\Delta\varepsilon_{\rm exch}}

\begin{document}

\title{The $SU(4)$ Kondo effect in double quantum dots with ferromagnetic leads}

\author{Ireneusz Weymann}
\email{weymann@amu.edu.pl}
\affiliation{Faculty of Physics, Adam Mickiewicz University, 
			 ul. Umultowska 85, 61-614 Pozna{\'n}, Poland}
			 
\author{Razvan Chirla}
\affiliation{Department  of  Physics, University  of  Oradea,  410087,  Oradea,  Romania}
\affiliation{Faculty of Medicine and Pharmacy, Department of Preclinical Sciences, University  of  Oradea, 410087, Oradea, Romania}

\author{Piotr Trocha}
\affiliation{Faculty of Physics, Adam Mickiewicz University, 
			 ul. Umultowska 85, 61-614 Pozna{\'n}, Poland}

\author{C\u at\u alin Pa\c scu Moca}
\affiliation{Department  of  Physics,  University  of  Oradea,  410087,  Oradea,  Romania}
\affiliation{BME-MTA  Exotic  Quantum  Phases  Research Group,   Budapest  University  of  Technology  and  Economics,  1521  Budapest,  Hungary}


\begin{abstract}
We investigate the spin-resolved transport properties, such as the linear
conductance and the tunnel magnetoresistance, of a double quantum dot device
attached to ferromagnetic leads and look for signatures of $SU(4)$ symmetry in the Kondo regime.
We show that the transport behavior greatly depends on the magnetic configuration of the device,
and the spin-$SU(2)$ as well as the orbital and spin-$SU(4)$ Kondo effects become generally
suppressed when the magnetic configuration of the leads varies from the 
antiparallel to the parallel one.
Furthermore, a finite spin polarization of the leads lifts the spin degeneracy
and drives the system from the $SU(4)$ to an orbital-$SU(2)$ Kondo state. We analyze 
in detail the crossover and show that the Kondo temperature between the two fixed points
has a non-monotonic dependence on the degree of spin polarization of the leads.
In terms of methods used, we characterize transport by using a combination of
analytical and numerical renormalization group approaches.
\end{abstract}

\maketitle


\section{Introduction}


Transport properties of double quantum dots (DQDs)
---the simplest realizations of artificial molecules \cite{blickPRB96}---
reveal a plethora of phenomena
not present in single quantum dot setups
\cite{gossardPRL95,hawrylakPRB95,zieglerPRB00,derWielRMP03,mcclurePRL07}.
In particular, in the regime of weak coupling
between DQD and external electrodes,
the interplay of Fermi statistics and charging effects
can result in the Pauli spin blockade 
effect~\cite{ono02,franssonPRB06,weymannPRB08}.
On the other hand, in the strong coupling regime,
the many-body electron correlations can result
in exotic Kondo effects~\cite{kondo64,NozieresJP80,hewson_book,CoxAP98,PustilnikPRL01},
such as the two-stage~\cite{PustilnikPRL01,vojtaPRB02,vanderWielPRL02,
Craig2004Science04,cornagliaPRB05,SasakiPRL09,ZitkoPRB10,tanakaPRB12,PetitPRB14}
or $SU(4)$ Kondo phenomena~\cite{bordaPRL03,SatoP05,GalpinPRL05,GalpinJPCM06,
AmashaPRL13,NishikawaPRB13,RuizPRB14,VernekAPL145,NishikawaPRB16}.
In the latter case, the ground state of the system
needs to exhibit a four-fold degeneracy,
which in the case of DQDs is assured by 
the spin and orbital degrees of freedom.
In fact, the presence of the $SU(4)$ Kondo effect
in double quantum dots has recently
been confirmed experimentally by A. Keller {\it et al.}~\cite{kellerNP14}.
By applying Zeeman and pseudo-Zeeman fields
to break the ground state degeneracy,
it was shown that the measured enhancement of the conductance
was indeed due to the formation of the $SU(4)$-symmetric Kondo state.

The emergence of the Kondo effect can however be hindered
by the presence of external perturbations or correlations in the leads.
In particular, when a quantum dot is attached to ferromagnetic electrodes, the Kondo effect
becomes affected due to the development of an exchange field $\exch$
induced by spin-dependent hybridization~\cite{martinekPRL03,
martinekPRL03NRG,pasupathy_04,hauptmannNatPhys08}.
Such an exchange field results in a splitting similar to the
Zeeman splitting in an external magnetic field~\cite{GaassPRL11},
still, its sign and magnitude can be tuned by a gate voltage
\cite{martinekPRB05,weymannPRB11,Csonka2012May}.
For single-level quantum dots, when the exchange field is getting larger than the corresponding
Kondo temperature $T_K$, the Kondo resonance starts to split.
The local density of states exhibits then 
only small satellite peaks at energies corresponding to $|\exch|$ 
\cite{pasupathy_04,hauptmannNatPhys08,GaassPRL11},
instead of a pronounced Abrikosov-Suhl resonance~\cite{hewson_book,goldhaber-gordon_98,Cronenwett98}.
For multi-dot structures, the transport behavior is generally more complex
and results from a subtle interplay of the relevant energy scales,
with the exchange field playing an important role~\cite{zitkoPRL12,wojcikPRB14}.

In this paper we investigate the linear conductance and the tunnel magnetoresistance 
in a double quantum dot device 
and analyze how transport is affected 
by the presence of ferromagnetic electrodes. We construct the full stability diagram,
and identify the regions where the spin-$SU(2)$, orbital-$SU(2)$ and the full $SU(4)$ 
Kondo states develop. The mere presence of the spin polarization in the leads lifts
the spin-degeneracy through the exchange field, which, at some particular points
in the stability diagram
drives the system through a crossover from an $SU(4)$ to an orbital-$SU(2)$ Kondo state \cite{Tosi}. 
We analyze this crossover in detail by using the scaling renormalization group (RG) 
approach~\cite{hewson_book}.
Furthermore, we investigate the effect of temperature on the 
linear conductance and identify ways to pinpoint the regions where Kondo states emerge 
by analyzing the system's behavior
in the two possible magnetic configurations of the leads (parallel or antiparallel).
Because an accurate analysis of such effects requires
resorting to nonperturbative methods,
here we employ the numerical renormalization group (NRG) method
\cite{WilsonRMP75,BullaRMP08}.

This paper is organized as follows:
In Sec. \ref{sec:Hamiltonian} we introduce the Hamiltonian
of the system under investigation. The
renormalization group analysis for the $SU(4)\to SU(2)$
crossover together with the scaling equations
that describe the crossover are presented in Sec. \ref{sec:SU4},
while Sec. \ref{sec:NRG} gives details on the NRG procedure
and presents how the quantities of interest, such as the linear conductance,
and computed for different magnetic configurations of the device.
Results of the NRG calculations for the $SU(4)\to SU(2)$ crossover are presented in Sec. \ref{sec:cross},
whereas the general behavior
of linear conductance and tunnel magnetoresistance is discussed in Sec. \ref{sec:stability}.
The paper is concluded in Sec. \ref{sec:conclusions}.


\section{Model for the double dot setup}
\label{sec:Hamiltonian}

\begin{figure}[!tb]
  \includegraphics[width=0.85\columnwidth]{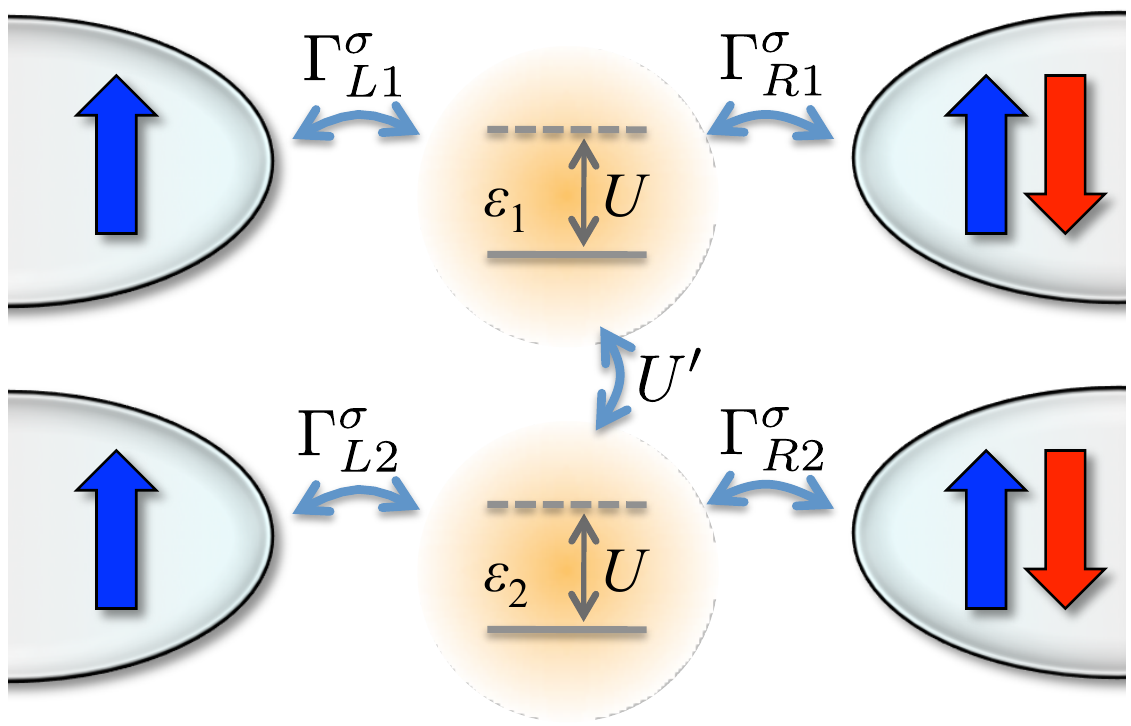}
  \caption{\label{fig:sketch}
  Schematic of a double quantum dot (DQD) system with ferromagnetic leads.
  Each dot, with energy level $\e_j$ and Coulomb correlation $U$,
  is coupled to a pair of left and right leads 
  with coupling strength $\Gamma_{rj}^\s$.
  The Coulomb correlations between the dots are denoted by $U'$.
  The magnetizations of the leads are assumed to form
  either a parallel (P) or antiparallel (AP) magnetic configuration.}
\end{figure}

The setup we consider consists of two capacitively coupled 
quantum dots, each one coupled to external leads (see the sketch in Fig.~\ref{fig:sketch}).
Each dot is described by the single impurity Anderson model (SIAM). We denote by 
$\varepsilon_{j}$, with $j=\{1,2\}$, the energy of an electron residing in dot $j$. Each 
dot can accommodate up to two electrons, and they interact with each other
through an on-site interaction $U$ and an interdot interaction $U'$. Their occupation
is denoted by $n_{j\sigma} = d^\dag_{j\sigma} d_{j\sigma}$,
with $d^\dag_{j\sigma}$ creating a spin-$\sigma$ electron in dot $j$.
The double dot Hamiltonian then reads
\begin{eqnarray}
  H_{\rm DQD} &=&
  \sum_{j\sigma} \varepsilon_{j} n_{j\sigma} + \sum_{j}  U
  n_{j\uparrow}n_{j\downarrow} \nonumber \\
  &+& U^\prime (n_{1\uparrow} + n_{1\downarrow})(n_{2\uparrow} + n_{2\downarrow})
 \,.\label{eq:H_DQD}
\end{eqnarray}
In the absence of an external magnetic field, $B=0$, if the energy levels are
degenerate, i.e. $\varepsilon_1=\varepsilon_2$, and when 
$U=U'$, the $H_{\rm DQD}$ Hamiltonian is $SU(4)$ invariant~\cite{hewson_yunori,fn1}. 
When the orbital degeneracy is lifted, corresponding to a situation when 
$\varepsilon_1\ne \varepsilon_2$, $H_{\rm DQD} $
remains $SU(2)$ invariant in the spin sector.
For more realistic situations \cite{kellerNP14}, when $U'/U<1$, the $SU(4)$
symmetry is in general lost.
Still, in this case, the system exhibits a special point in the $\{\varepsilon_1,\varepsilon_2\}$
parameter space where an emergent $SU(4)$ symmetry can occur~\cite{hewson_yunori}, i.e. 
$\{ \varepsilon_1,\varepsilon_2 \} \approx \{-U'/2, -U'/2\}$~\cite{fn2}.
This special point will be discussed in more detail in Secs.~\ref{sec:SU4} 
and \ref{sec:cross}.
The double dot setup is attached to four external ferromagnetic leads,
modeled as reservoirs of noninteracting quasiparticles,
\begin{equation}
  H_{\rm Leads} = \sum_{rj{\bf k}\sigma} \e_{rj{\bf k}\sigma}
  c^{\dagger}_{rj{\bf k}\sigma} c_{rj{\bf k}\sigma}\,.\label{eq:H_Leads}
\end{equation}
Here, $c^{\dagger}_{rj{\bf k}\sigma}$ is the
creation operator for an electron with momentum ${\bf k}$ and spin $\sigma$
in the lead $r=\{L,R\}$ 
attached to  dot $j$. Consequently, the corresponding local density of states
$\rho_{rj}^\sigma$ becomes spin dependent. Furthermore, this affects the
broadening function that describes the coupling between the dots and the leads, i.e.
$\Gamma_{rj}^\sigma = \pi \rho_{rj}^\sigma |v_{rj}|^2$, where $v_{rj}$ is the 
amplitude of the tunneling. The tunneling Hamiltonian is given by
\begin{eqnarray}
  H_{\rm Tun} = \sum_{rj {\bf k} \sigma}
  v_{rj} \left( c^\dagger_{rj{\bf k}\sigma} d_{j \sigma} + d^\dagger_{j \sigma} c_{rj{\bf k}\sigma}  \right) .\label{eq:H_Tun}
\end{eqnarray}
It is more convenient to express the couplings in terms
of spin polarization of a given lead, $p_{rj}$, as
$\Gamma_{rj}^\s = (1+\s p_{rj})\Gamma_{rj}$,
where $\Gamma_{rj} = (\Gamma_{rj}^\uparrow + \Gamma_{rj}^\downarrow) / 2$.
In the present work, we assume that the magnetizations of the leads are
collinear and can take two configurations: ($i$) parallel (P) and ($ii$) antiparallel (AP).
We also consider that the density of states is flat
with the bandwidth given by $2D$, and set $D\equiv 1$ as the energy unit.
The total Hamiltonian describing the double dot system coupled to ferromagnetic leads 
is then given by
\begin{equation}
H = H_{\rm DQD} + H_{\rm Leads} + H_{\rm Tun}.\label{eq:H}
\end{equation}
In the following we will solve it using the Wilson's NRG method~\cite{WilsonRMP75}. 

\section{The $SU(4)$ to $SU(2)$ crossover in the Kondo regime}\label{sec:SU4}

We shall first focus on the special point $\{\varepsilon_1, \varepsilon_2\} = \{-U'/2, U'/2\}$ that
displays the emerging $SU(4)$ physics (provided $U\gtrsim U'$~\cite{hewson_yunori})
in the limit when the leads are nonmagnetic. For finite spin
polarization of the leads, the spin degeneracy is lifted, but the orbital $SU(2)$ symmetry is 
preserved. So, by changing the polarization of the external leads, it is possible to capture the 
$SU(4)\to SU(2)$ crossover. To comprehend the essential physics we map the Hamiltonian~\eqref{eq:H} to 
the Kondo model by projecting onto the subspace with single occupancy $\langle n\rangle \simeq 1$
by using the Schrieffer-Wolff transformation~\cite{hewson_book}.  
We assume that the dots are symmetrically coupled, $v_{Lj}=v_{Rj}=v_j$ and $p_{Lj}=p_{Rj}=p$.
We then make a change of basis
by performing a unitary transformation on the leads operators and 
use an even/odd combination,
\begin{eqnarray}\label{eq:unitary}
\left(
\begin{array}{cc}
c_{e j\bf k\sigma} \\
c_{o j\bf k\sigma}
\end{array}
\right)&=&\frac{1}{\sqrt{2}}
\left(
\begin{array}{cc}
1 & 1 \\
-1  & 1
\end{array}
\right)
\left(
\begin{array}{cc}
c_{L j\bf k\sigma}  \\
c_{R j\bf k\sigma}
\end{array}
\right)\; .
\end{eqnarray}
In this even-odd  basis, the odd channel becomes decoupled and the
double-dot remains coupled only to the even channel.
In what follows, we shall drop the corresponding subscript, i.e.,
$c_{e j\bf k\sigma}\to c_{j\bf k\sigma}$.
We introduce the tensor product notations 
$(\hat{\sigma}^{\mu}\otimes\hat{\tau}^{\nu})_{ j\sigma ; j' \sigma'}
=\sum_{\textbf{kk}'}c_{j\textbf{k}\sigma}^{\dagger} \sigma^{\mu}_{\sigma \sigma'}\tau^{\nu}_{jj'}
c_{j'\textbf{k}'\sigma'}$ and  $(\hat{S}^{\mu}\otimes\hat{T}^{\nu})_{j \sigma ; j' \sigma' }=d_{j\sigma}
^{\dagger}\, (\frac{1}{2}\sigma^{\mu}_{\sigma \sigma'})(\frac{1}{2}\tau^{\nu}_{jj'})\,d_{j'\sigma'}$,
where $\sigma^{\mu}=\{I_2, \sigma_x, \sigma_y, \sigma_z \}$ are the regular
Pauli matrices for $\mu=1\to 3$ and the unit matrix
when $\mu=0$, acting in the spin degrees of freedom, and similar for 
$\tau^{\nu}$ but acting on the orbital part. Then, 
disregarding the potential scattering, the anisotropic Kondo Hamiltonian can be written as
\begin{equation}
H_{\mathrm{K}}=
\sum_{\substack{\sigma\sigma'\alpha \alpha'\\ jj'ii'\\ \mu\nu }}
 J^{\mu;\nu}_{j \sigma ;j' \sigma' }(\hat{\sigma}^{\mu}\otimes\hat{\tau}^{\nu})_{j \sigma ; j' \sigma' }(\hat{S}^{\mu}\otimes\hat{T}^{\nu})_{i \alpha ;i' \alpha'}\;.\label{eq:H_K}
\end{equation}
%
%
Altogether there are 15 terms in Eq.~\eqref{eq:H_K}
and the exchange couplings $J^{\mu;\nu}_{j \sigma ;j' \sigma' }$
depend on all the parameters of the original Hamiltonian,
i.e. $\varepsilon_j, U$ and $\Gamma_{rj}^\sigma$.
In the limiting case when $U=U'$ and $p=0$, it is straightforward to show
that all the couplings are equal $\mathbf J\to J$ and the charge and spin contributions 
combine in an $SU(4)$-symmetric way. The $15=4^2-1$ generators for the $SU(4)$ Lie algebra are 
$\{I_2, \boldsymbol \sigma\}\otimes \{I_2,\boldsymbol \tau\}-I_2\otimes I_2$. On the other hand, 
when $p=1$, i.e. the leads are frozen for example in the spin-$\uparrow$ state, $H_{\rm K}$ remains
$SU(2)$ invariant in the orbital sector.

\begin{figure}[!tb]
\includegraphics[width=0.95\columnwidth]{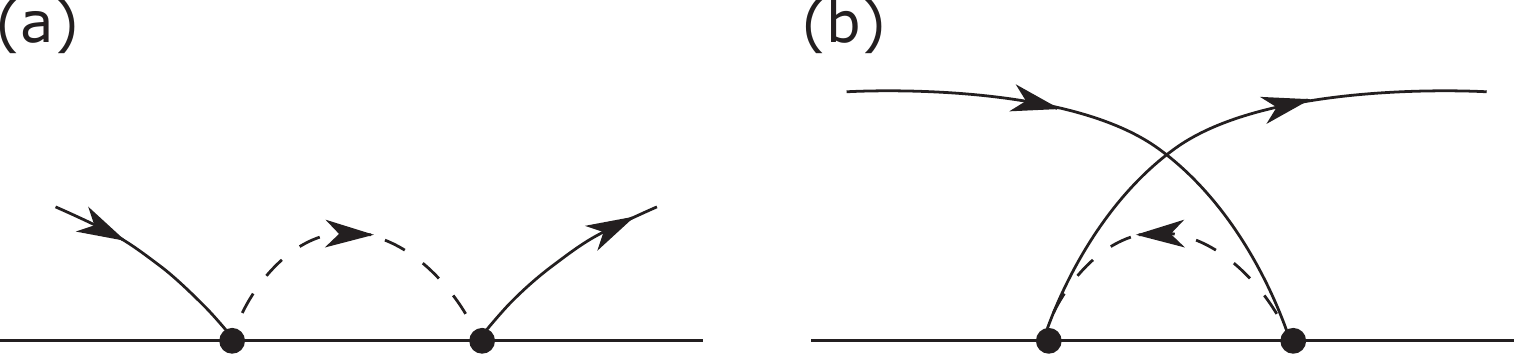}
\caption{Second order diagrams contributing to  
the renormalization of the coupling matrix displaying processes when a virtual particle
is scattered in the upper band edge (a) or a virtual hole in the lower band edge (b) of the lead electrons.
} 
\label{fig:rg}
\end{figure}

\begin{figure}[!tb]
\includegraphics[width=0.9\columnwidth]{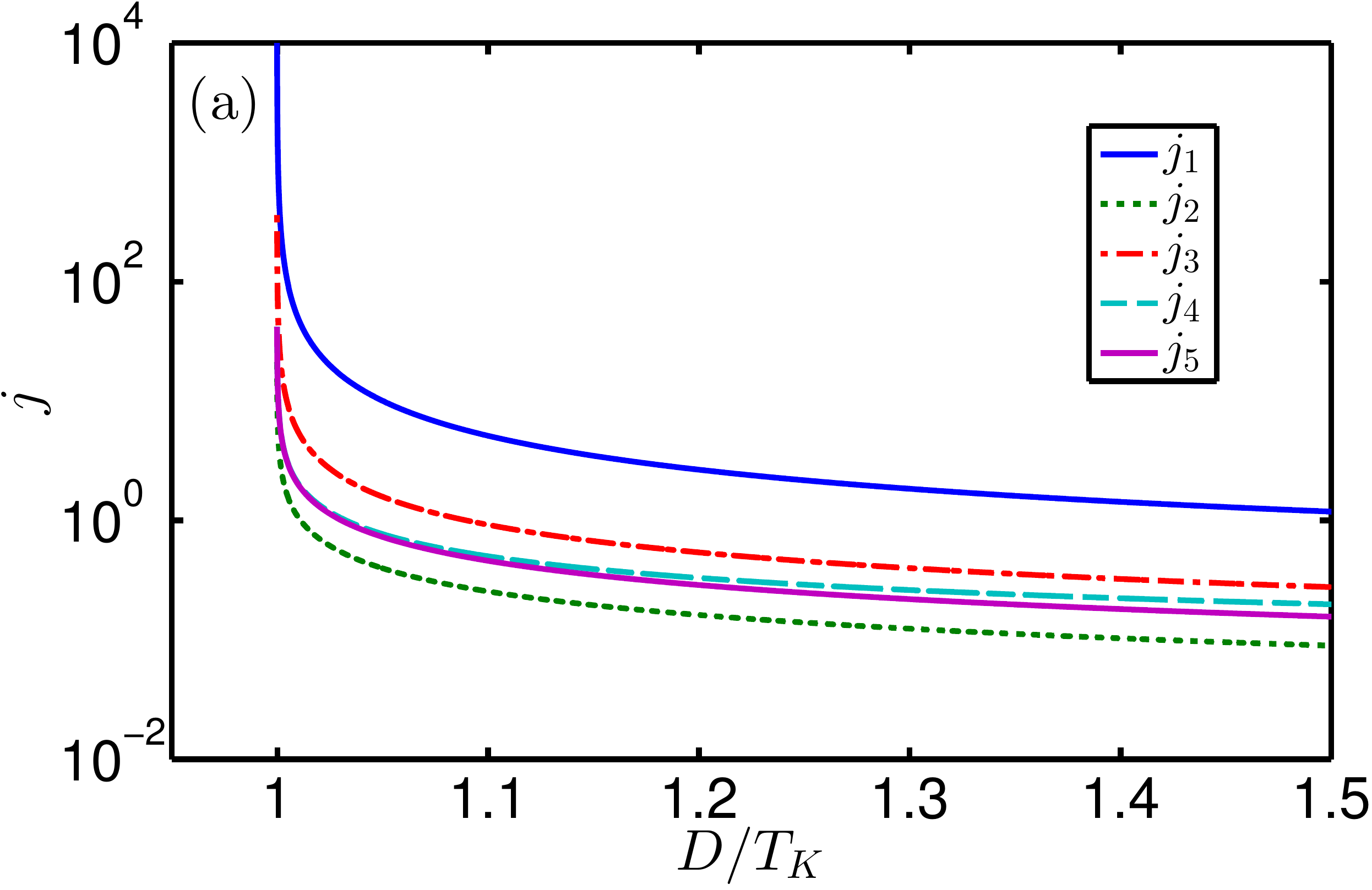}
\includegraphics[width=0.88\columnwidth]{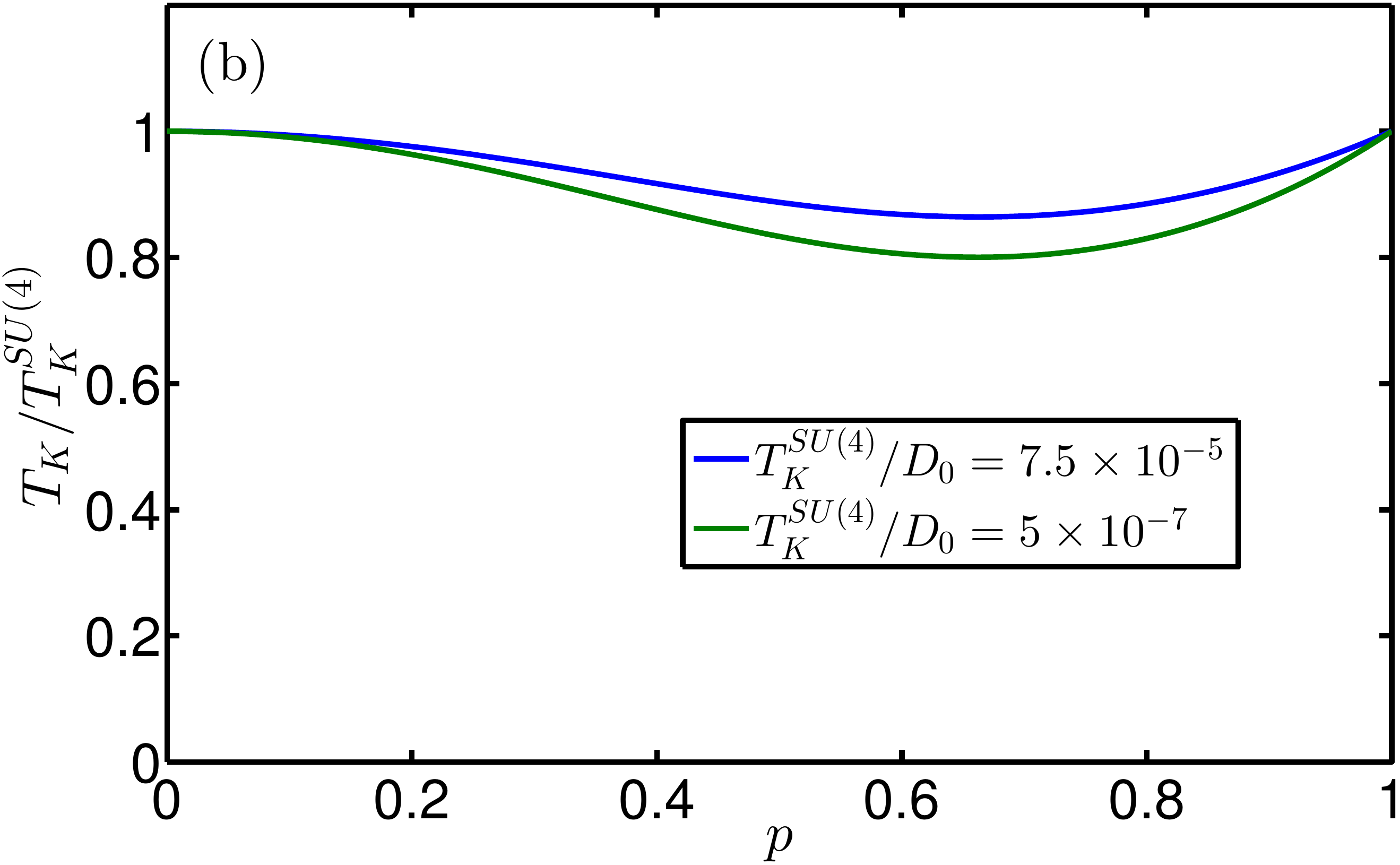}
\caption{(a) The renormalization of the coupling constants
as the bandwidth is changed.
We used $p=0.8$ and $\rho_0 J^0=0.026$.
For this choice of parameters, $T_K/D_0=6.6 \times 10^{-5}$.
(b) The evolution of $T_K$ as a function of spin polarization $p$. 
} 
\label{fig:j_scaling}
\end{figure}

To capture the crossover we performed the RG analysis~\cite{hewson_book}
for the exchange couplings $\mathbf J$ in between these 
two fixed points. The second-order processes  (particle and hole-like)
that renormalize the couplings are displayed in Fig.~\ref{fig:rg}. Keeping in mind 
that the polarization of the leads affects only the spin sector,
we can group the couplings into $5$ distinct classes.
Furthermore, we define dimensionless couplings
by introducing the local density of states $\rho_0=1/2D_0$ as
\begin{eqnarray}
j_1=&\rho_0 J^{\mu=\{0,3\};\nu\neq 0}_{j \uparrow;j'\uparrow}\;,
&j_2=\rho_0 J^{\mu=\{0,3\};\nu\neq 0}_{j \downarrow; j' \downarrow } ,\nonumber \\
j_3=&\rho_0 J^{\mu=\{1,2\};\nu}_{j \sigma; j' \overline{\sigma}}\;,
&j_4=\rho_0 J^{3;0}_{j\uparrow; j \uparrow }\, ,\\
&j_5=\rho_0 J^{3;0}_{j\downarrow; j \downarrow}\, ,  \nonumber
\label{eq:couplings}
\end{eqnarray}
subject to initial conditions $j^0_1=j^0_4=\rho_0 J^0 (1+p), 
j^0_2=j^0_5=\rho_0 J^0 (1-p)$ and $j^0_3=\rho_0 J^0 \sqrt{1-p^2}$, 
where $J^0=v^2\left (\frac{1}{\varepsilon+U}-\frac{1}{\varepsilon}\right )$ 
and $v$ is the isotropic coupling~\cite{fn3}. Here $2D_0$ is the bandwith for the conduction electrons. 
To second order in $\mathbf j$, the scaling equations are easily derived 
by progressively reducing the bandwidth $D$~\cite{hewson_book} as
\begin{eqnarray}\label{eq:rg_eq}
&&\frac{dj_1}{d \ln D}=-2j_3^2-2j_1^2\;\nonumber\\
&&\frac{dj_2}{d \ln D}=-2j_3^2-2j_2^2\;\nonumber\\
&&\frac{dj_3}{d \ln D}=-\frac{3}{2}j_3(j_1+j_2)-\frac{1}{2}j_3(j_4+j_5)\;\\
&&\frac{dj_4}{d \ln D}=-4j_3^2\;\nonumber\\
&&\frac{dj_5}{d \ln D}=-4j_3^2\;\nonumber.
\end{eqnarray}
The $SU(4)$ fixed point is captured by setting $p=0$, in which case 
the set~\eqref{eq:rg_eq} of equations collapses to a single one, i.e.
$dj/d \ln D=-4j^2$. 
 In contrast, when the leads are fully polarized, $p=1$, 
the coupling $j_4$ remains marginal while $j_1$ rescales 
accordingly to the regular $SU(2)$ Kondo physics,
$dj_1/d \ln D=-2j_1^2$. In the general situation
we can solve the RG equations~\eqref{eq:rg_eq}
numerically. A typical solution is presented
in Fig. \ref{fig:j_scaling}(a) for $p=0.8$, and as expected
all the couplings diverge at the same characteristic energy
scale that can be associated with the Kondo temperature~\cite{hewson_book}.

In general, in the $SU(N)$ Kondo model~\cite{hewson_book}, apart from
some higher-order corrections~\cite{moca_mora},
the Kondo temperature is  $T_K^{SU(N)}\simeq D_0\, e^{-1/(N j_0)}$. On the other hand, when the polarization of the leads
is changed from $p=0\to 1$, we double the exchange interaction, so that $T_K$ is expected to remain 
unchanged. To test this conjecture, 
we represent in Fig.~\ref{fig:j_scaling}(b) the evolution of $T_K$ with 
the spin polarization $p$ of the leads, which indeed shows that 
$T_K$ is the same at the two fixed points.
When $U \gg U'$, depending on the ratios $\Gamma/U$ and $\Gamma/U'$,
the two characteristic energy scales, $T_K^{SU(2)}$  and $T_K^{SU(4)}$, can be well separated, 
but otherwise the physics remains the same.

To conclude this section, 
the set~\eqref{eq:rg_eq} of RG equations describes consistently the  
$SU(4)\to SU(2)$ crossover  and captures the essential Kondo physics in between the two fixed points. 
In Sec.~\ref{sec:cross}, we supplement the RG analysis with more exact numerical renormalization 
group calculations~\cite{WilsonRMP75,FlexibleDMNRG} and focus on computing
measurable quantities such as the conductance and the tunnel magnetoresistance.

\section{Numerical renormalization group and the conductance}
\label{sec:NRG}

In this work we are interested in the linear response transport
properties of the system at low enough temperatures such that the
electron correlations give rise to the Kondo effect
\cite{kondo64,hewson_book}.
The aim, in particular, is to elucidate the role
of spin-dependent tunneling on the transport properties
in the full parameter space, with a special focus on the $SU(4)$
Kondo regime \cite{bordaPRL03}.
In order to achieve this goal in the most accurate manner,
we employ the nonperturbative numerical renormalization group (NRG) method
\cite{WilsonRMP75,FlexibleDMNRG}.
In the NRG approach, the conduction bands of the non-interacting electrons in the leads
are discretized in a logarithmic way
with a discretization parameter $\Lambda$ (here we use $\Lambda=2$).
The discretized Hamiltonian is then transformed to a tight-binding chain Hamiltonian
with exponentially decaying hoppings (Wilson chain).

We follow the same strategy as discussed in Sec. \ref{sec:SU4}
and use the even-odd basis. In this way each dot is coupled 
to a single channel -- the even channel --  with a 
coupling strength, $\Gamma_j^\s = \Gamma_{Lj}^\s + \Gamma_{Rj}^\s$.
The NRG Hamiltonian of the system is 
\begin{eqnarray}
H_{\rm NRG} &=& H_{\rm DQD} + \sum_{j\s}\sqrt{\frac{\Gamma_{j}^\s}{\rho_0 \pi}} \left(f_{j0\s}^\dag d_{j\s} + d_{j\s}^\dag f_{j0\s} \right) \nonumber\\
&+& \sum_{jn \s} \xi_n \left( f^\dag_{jn\s} f_{jn+1\s} + f_{jn+1\s}^\dag f_{jn\s}  \right).
\end{eqnarray}
Here, $f^\dag_{jn\s}$ denotes the creation operator
of a spin-$\sigma$ electron at site $n$ ($n=0,1,2\dots$) of the $j^{\rm th}$ ($j=1,2$) Wilson chain
and $\xi_n$ are the respective hopping integrals.
This Hamiltonian is solved iteratively
by retaining an appropriate number $N_K$ of low-energy states 
at each iteration (here we keep at least $N_K = 10^4$ states).
The discarded states, on the other hand,
form a complete many-body basis of the whole NRG Hamiltonian
\cite{AndersSchiller05}
and are used to construct the full density matrix of the system
\cite{WeichselbaumPRL07}.

Along the NRG procedure,
one needs to deal with a large Hilbert space at each step 
of iteration, therefore it is  crucial to exploit
as many symmetries of the NRG Hamiltonian as possible.
Here we make use of four Abelian symmetries~\cite{fn4}, defined by the generators
\begin{eqnarray}
Q_j &=& \sum_\sigma \left(n_{j\sigma}  - \tfrac{1}{2}\right)
+ \sum_{n\s}\left( f_{jn\sigma}^\dag f_{jn\sigma} - \tfrac{1}{2} \right), \\ 
S_{z}^j &=& \frac{1}{2}\left(n_{j\uparrow} - n_{j\downarrow}\right)
+ \frac{1}{2}\sum_{n}\left( f_{jn\uparrow}^\dag f_{jn\uparrow} - f_{jn\downarrow}^\dag f_{jn\downarrow} \right),\nonumber
\end{eqnarray}
for the total charge and $z^{\rm th}$ spin component of dot and chain $j$, respectively.
The quantities we are particularly interested in
are $(i)$ the total spectral function 
\begin{equation}
A(\omega) = \sum_{j\s} A_{j\s}(\omega) =  -\frac{1}{\pi} \sum_{j\s} {\rm Im} G_{j\s}^R(\omega),
\end{equation}
with $G_{j\s}^R(\omega)$ being the Fourier transform of the retarded
Green's function, $G_{j\s}^R(t) = -i\Theta(t) \langle \{d_{j\s}(t),d_{j\s}^\dag(0) \} \rangle$,
and $(ii)$ the linear conductance
\begin{equation}\label{eq:conductance}
G = 
\frac{e^2}{h} \sum_{j\s} \frac{4\Gamma_{Lj}^\sigma\Gamma_{Rj}^\sigma}
{\Gamma_{Lj}^\sigma+\Gamma_{Rj}^\sigma} \int \!\! d\omega \left(\!\!-\frac{\partial f(\omega) }{\partial \omega} \right)  \pi A_{j\sigma}(\omega),
\end{equation}
where $f (\omega)$ denotes the Fermi-Dirac distribution function~\cite{fn5,MeirPRL92}.
To get a clear picture, we assume equal spin polarizations of the leads,
$p_{rj} \equiv p$, and equal coupling strengths,
$\Gamma_{rj} \equiv \Gamma/2$. Then the 
expression~\eqref{eq:conductance}
for the
 linear conductance reduces to
\begin{equation}\label{Eq:GAP}
G^{\rm AP} = 
\frac{2e^2}{h} (1-p^2)\Gamma \int \!\! d\omega \left(\!\! -\frac{\partial f (\omega)}{\partial \omega} \right)  \pi A^{\rm AP} (\omega),
\end{equation}
for the antiparallel (AP) configuration, with $A_{j}^{\rm AP} (\omega) = A_{j\uparrow}^{\rm AP} (\omega)
= A_{j\downarrow}^{\rm AP} (\omega)$ the spectral function in the AP configuration.
As can be seen, $G^{\rm AP}$ is the linear conductance 
-- up to the prefactor $(1-p^2)$ --
of a DQD setup with nonmagnetic leads.
On the other hand, the conductance 
in the parallel (P) configuration is given by
\begin{equation}
G^{\rm P} = 
\frac{e^2}{h} \sum_\sigma (1+\sigma p) \Gamma \int \!\! d\omega \left(\!\! -\frac{\partial f (\omega)}{\partial \omega} \right)  \pi A_{\sigma}^{\rm P} (\omega),
\end{equation}
where $A_{j\sigma}^{\rm P} (\omega)$ is the  spectral function in the parallel configuration.
The difference between the conductances in both magnetic alignments
can be described by the tunnel magnetoresistance, which is defined as,
\cite{julliere,barnasJPCM08}
${\rm TMR} = G^{\rm P} /  G^{\rm AP} - 1$.
In the present work we use the NRG to investigate
the full phase space of the model. However, to connect to the RG results presented in 
Sec. \ref{sec:SU4}, let us first discuss the 
$SU(4) \to SU(2)$ crossover and follow the evolution of the 
spectral functions as well as of the conductance, quantities that were not 
accessible in the RG approach. 

\section{The $SU(4)$ to $SU(2)$ crossover: NRG results}\label{sec:cross}

\begin{figure}[!tb]
  \includegraphics[width=0.9\columnwidth]{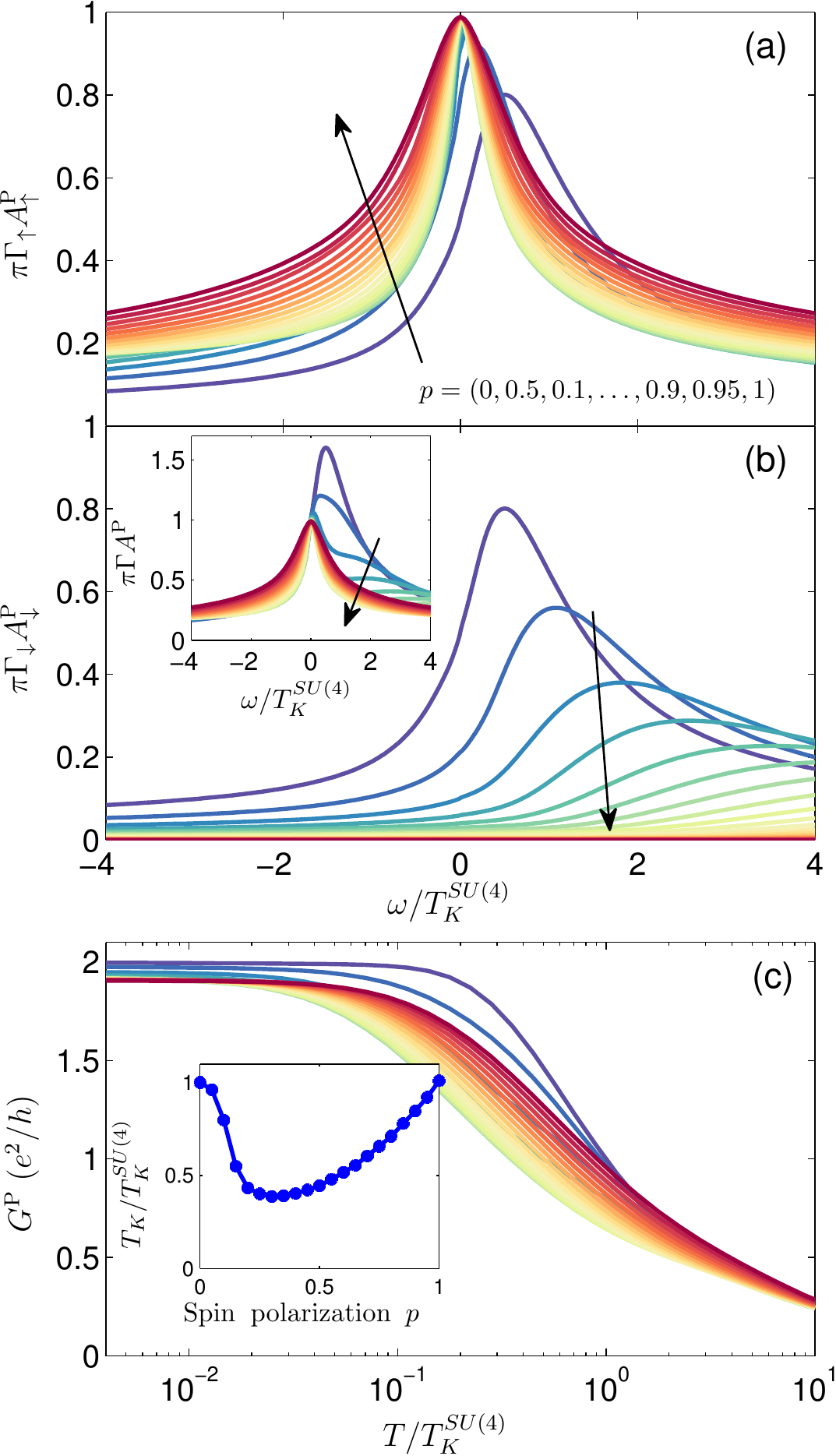}
  \caption{\label{Fig:spectral_SU4}
  The energy dependence of (a) the zero-temperature spin-up and (b) spin-down spectral function,
  together with (c) the linear conductance as a function of temperature
  calculated for different spin polarization of the leads,
  ranging from $p=0$ to $p=1$ in steps of 0.05
  (the arrow indicates the direction in which $p$ increases),
  in the $SU(4)$ Kondo regime.
  The inset in (b) shows the total spectral function,
  while the inset in (c) presents the Kondo temperature 
  as a function of $p$. The Kondo temperature
  is defined by $G(T)/G(T=0) = 1/2$. $T_{K}^{SU(4)}$ ($\approx 2.8\cdot 10^{-4}U$)
  denotes the $SU(4)$ Kondo temperature (in the case of $p=0$).
  The parameters are: $U=U'=0.5$, $\Gamma = 0.015$,
  in units of band halfwidth, and $\e_1 = \e_2 = -U'/2$.}
\end{figure}

\begin{figure}[!tb]
  \includegraphics[width=0.9\columnwidth]{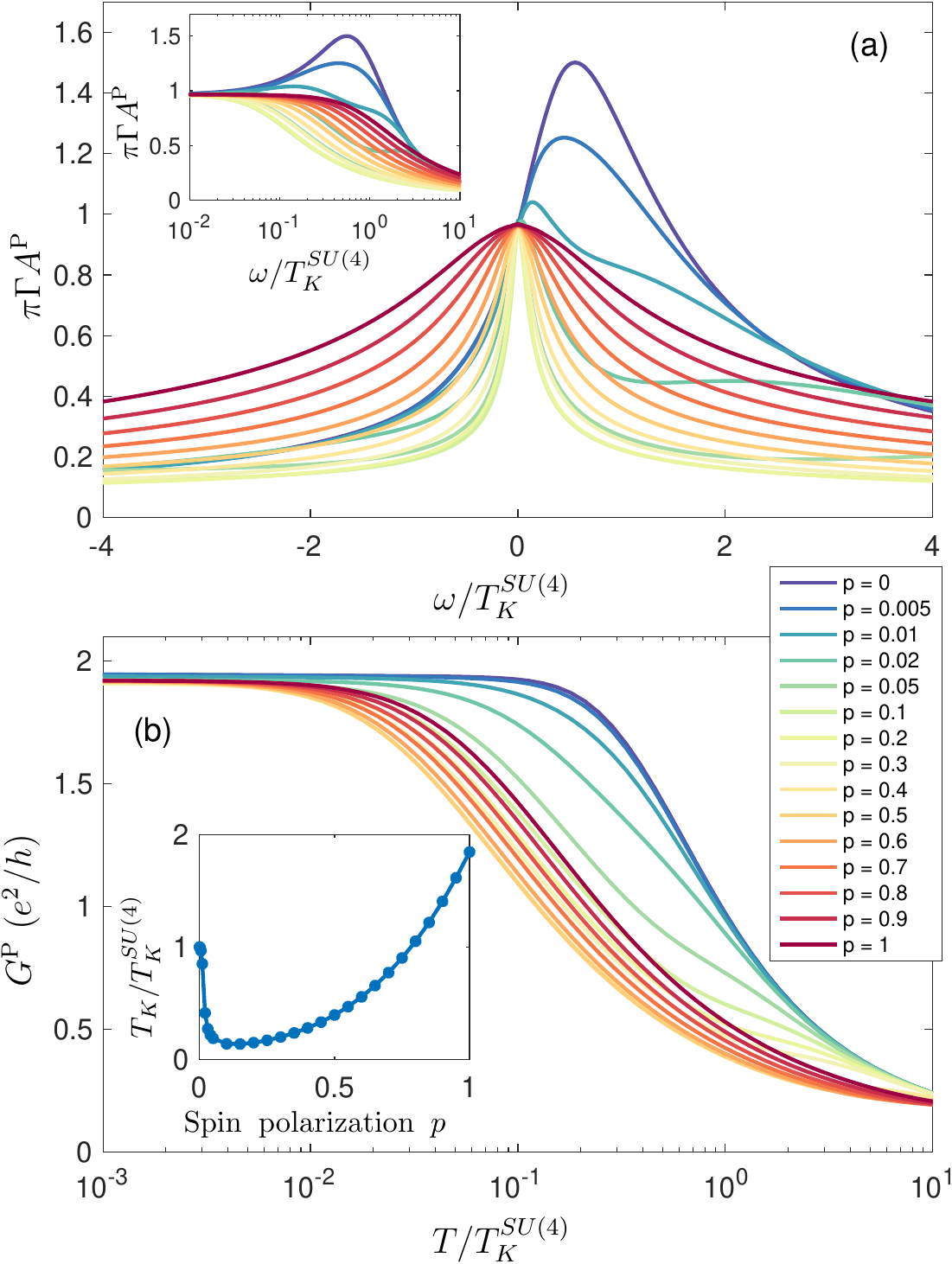}
  \caption{\label{Fig:spectral_SU4_2}
  (a) The energy dependence of the zero-temperature total spectral function,
  and (b) the temperature dependence of the linear conductance
  calculated for different spin polarization of the leads, as indicated.
  The inset in (a) shows the spectral function on the logarithmic scale,
  while the inset in (b) presents the Kondo temperature 
  as a function of $p$. 
  The parameters are the same as in \fig{Fig:spectral_SU4} with $U=1$ and $U'=U/2$.
  Now $T_{K}^{SU(4)}/U\approx 7.5\cdot 10^{-5}$.}
\end{figure}

In this section we focus on the $SU(4)$ Kondo regime
and analyze the influence of finite leads' spin polarization
on the transport properties. We shall present the
results for the spectral functions $A_{\sigma}^{\rm P} (\omega)$ 
as well as for the temperature dependence of the 
conductance~\cite{fn6}. We will discuss in detail the 
case of $U=U'$, and later address a more realistic situation when $U > U'$.

An important quantity that captures the
crossover is the normalized spectral function, $A_{\sigma}^{\rm P} (\omega)$, 
whose spin components are displayed in  Figs.~\ref{Fig:spectral_SU4}(a) and (b), respectively. 
The total spectral function itself, $A^{\rm P} (\omega)$, is presented in the 
inset of Fig.~\ref{Fig:spectral_SU4}(b). 
When $p=0$, it displays the regular $SU(4)$ Kondo resonance formed
away from the Fermi level at $\omega \approx T_K^{SU(4)}$.
When increasing the spin polarization 
its maximum becomes suppressed and moves
toward $\omega=0$, and when $p=1$
the orbital-$SU(2)$ Kondo resonance is formed at $\omega =0$.

We can get more information by inspecting the spin-resolved
spectral functions.
In the case of spin-up channel, which  belongs to the majority-spin subband,
increasing the spin polarization results in an enhancement
of the spectral function to $A_\uparrow^{\rm P}(\omega\to 0) \simeq 1/\pi\Gamma_\uparrow$.
Moreover, the maximum in $A_\uparrow^{\rm P}(\omega)$
gradually shifts to the Fermi energy,
such that for $p=1$, only the orbital degree of freedom is relevant,
and the $SU(2)$ Kondo peak becomes symmetric around $\omega=0$.
On the other hand, $A_\downarrow^{\rm P}(\omega)$
exhibits a completely different behavior.
First of all, increasing the spin polarization results
in a decrease of $A_\downarrow^{\rm P}(\omega)$.
Furthermore, the maximum in the spin-down spectral function
moves away from the Fermi energy, 
due to the development of the exchange field
 $\Delta \varepsilon_{\rm exch}$~\cite{martinekPRL03,
martinekPRL03NRG} and this splitting grows with increasing $p$.
Finally, for $p=1$, $A_\downarrow^{\rm P}(\omega)$
becomes completely quenched at low energies.

This distinct behavior of the spectral function
is corroborated with a detailed analysis of
the temperature dependence of the linear conductance,
which is shown in \fig{Fig:spectral_SU4}(c). At the two fixed points (corresponding to 
$p=0$ and $p=1$),
the conductance is a universal function
of $T/T_{K}^{SU(N)}$~\cite{bordaPRL03,kellerNP14}.

Interestingly, despite the fact that 
the system's ground state degeneracy becomes reduced from four-fold to two-fold,
increasing the spin polarization has a rather small effect on the conductance itself. 
Its temperature dependence allows us to define the 
Kondo scale as  $G^{\rm P}(T=T_K)=G^{\rm P}(T=0)/2$.  The evolution of $T_K$ 
with increasing the spin polarization is presented in the inset of Fig.~\ref{Fig:spectral_SU4}(c).
As previously predicted by the RG equations, the polarization of the leads 
has a relatively small effect on $T_K$ and, consequently,
$T_{K}^{SU(4)}\approx T_{K}^{SU(2)}$.
We would however like to note that the difference between the two 
Kondo temperatures can be enlarged by reducing the charge fluctuations,
i.e. by decreasing the ratio of $\Gamma / U$.

Let us now analyze a more realistic situation when $U> U'$.
Now the two Kondo temperatures $T_{K}^{SU(N=2,4)}$ are well separated, 
which allows us to clearly identify the exchange-field-induced splitting
in the conductance behavior. 
This can be obtained by properly tuning the ratio between the couplings
and Coulomb correlations. The energy dependence of the spectral function
and the temperature dependence of the conductance calculated for 
$\Gamma/U=0.015$ are shown in \fig{Fig:spectral_SU4_2}.
Since $T_{K}^{SU(4)}$ is now much smaller ($T_{K}^{SU(4)}/U\approx 7.5\cdot 10^{-5}$),
a very small spin polarization ($p\gtrsim 0.02$) is sufficient to suppress the $SU(4)$
Kondo effect completely [see \fig{Fig:spectral_SU4_2}(a)].
Quite unexpectedly, the width of the orbital Kondo peak
depends in a nonmonotonic fashion on the degree of spin polarization
of the leads [see also the inset in \fig{Fig:spectral_SU4_2}(a)],
and the minimum width occurs around $p \approx 0.1$.

This behavior is now clearly reflected in the temperature dependence of the conductance
shown in \fig{Fig:spectral_SU4_2}(b). The $p=0$ curve 
presents a universal $SU(4)$ conductance dependence,
which then, with increasing $p$, smoothly changes to the
$SU(2)$ universal curve. Moreover, the extracted Kondo temperature
reveals a nonmonotonic dependence on spin polarization.
First, the Kondo temperature quickly drops with $p$
and is much lower than $T_{K}^{SU(4)}$. Further
increase of $p$, however, results in an enhancement
of the $SU(2)$ Kondo temperature.
To understand this enhancement, we recall that
spin-dependent hybridization (which grows with $p$),
results in DQD level renormalization,
such that the position of the spin-up levels
becomes effectively lowered. As a consequence,
it reduces the excitation energies for the pseudo-spin-flip
processes responsible for the Kondo effect, leading to an increase of 
$T_{K}^{SU(2)}$, such that for $p=1$, one may even achieve
$T_{K}^{SU(2)} > T_{K}^{SU(4)}$, see the inset of \fig{Fig:spectral_SU4_2}(b),
which is not in general obvious.


\section{Stability diagrams and tunnel magnetoresistance }
\label{sec:stability}


\begin{figure}[t!]
  \includegraphics[width=0.95\columnwidth]{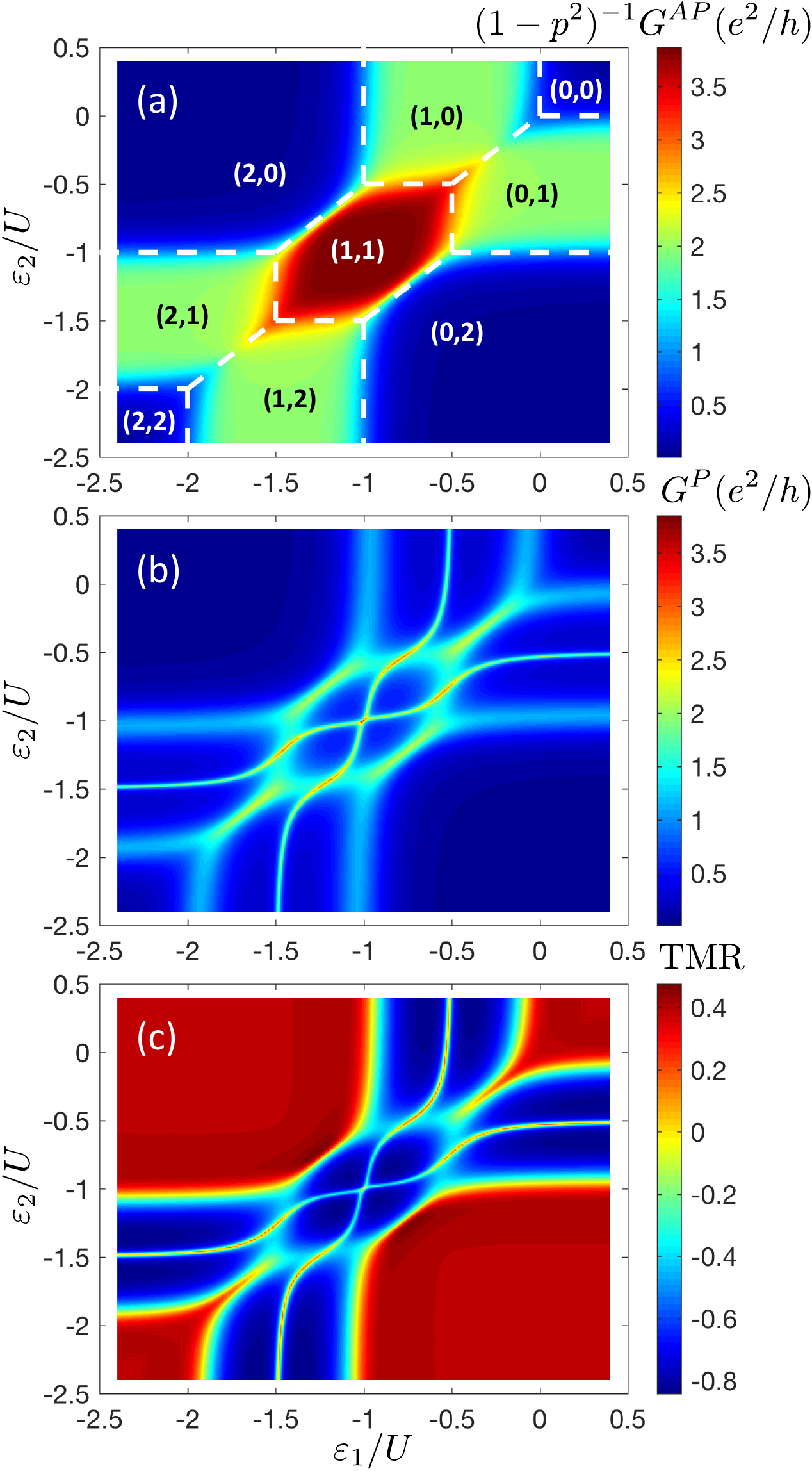}
  \caption{\label{Fig:stability}
  The linear conductance in (a) the antiparallel and (b) parallel magnetic configuration
  and (c) the resulting TMR as a function of DQD energy levels
  $\e_1$ and $\e_2$.
  The dashed lines in (a) mark the regions where the
  DQD is in a state ($n_1$,$n_2$) with $n_1$ ($n_2$) electrons in first (second) dot.
  The parameters are: $U=1$, $U'=0.5$, $\Gamma = 0.07$, $p=0.4$ and $T=10^{-6}$.}
\end{figure}

In this section we present results for the low-temperature linear 
conductance in the parallel and antiparallel configurations,
together with the TMR, calculated as a function of the double dot energy levels
$\e_1$ and $\e_2$. In Fig.~\ref{Fig:stability} we present a typical stability diagram
that covers the full parameter space, from empty to fully occupied DQD. 
In this section we address only the regime where  $U/U'=2$. 

Let us first discuss the case of the antiparallel magnetic configuration
shown in \fig{Fig:stability}(a).
The conductance shows a pattern that closely resembles that of nonmagnetic DQD system~\cite{LeoKouwenhoven}.
The dashed lines separate the equilibrium charged transport domains.
When the number of electrons in each dot is even, the DQD is in a singlet state,
no Kondo effect develops and the observed low conductance
results only from cotunneling processes.
However, when the electron number in either quantum dot is odd, 
the electronic correlations can give rise to an enhanced conductance due to the Kondo effect, 
provided the temperature is lower than the Kondo temperature.
In our calculations the assumed temperature is very low, $T\simeq 10^{-6} U$,
such that in each Coulomb blockade region the Kondo effect develops.

As the parameter space is relative large, 
depending on the nature of the ground state, several  
types of the Kondo effects develop.
When the occupancy of one of the dots is odd,
a typical spin-$SU(2)$ Kondo effect develops.
This can be observed in transport regime with
the electron numbers belonging to the set $\{(1,0), (0,1),
(2,1), (1,2)\}$ [see \fig{Fig:stability}(a)],
where $G^{\rm AP}/(1-p^2)$ reaches the unitary limit $\approx 2e^2/h$.
Since there is no direct hopping between the dots,
when every dot is singly occupied, $(n_1, n_2)= (1,1)$
one finds that the $SU(2)$ Kondo effect 
develops independently in each quantum dot, such that the 
total conductance reaches $G^{\rm AP}/(1-p^2) \approx  4e^2/h$.

The stability diagram allows us to get a better understanding of
how the emergent $SU(4)$ Kondo 
effect develops: along the line separating
the charge states $(0,1)\leftrightarrow (1,0)$, and 
$(2,1)\leftrightarrow (1,2)$, besides the spin degeneracy
an additional orbital degeneracy is present and the ground state
is four-fold degenerate. Consequently, the system exhibits the $SU(4)$ Kondo effect~\cite{kellerNP14}.
As we have seen in Sec.~\ref{sec:cross}, the $SU(4)$ Kondo state is better  
revealed in the parallel configuration,
where the spin degeneracy is broken.

The conductance in the parallel configuration is presented in \fig{Fig:stability}(b)
and reveals some huge differences when compared to the AP configuration. 
This is due to the emergence of the exchange field $\Delta \varepsilon_{\rm exch}$
that splits the levels of the DQD and lifts the spin degeneracy \cite{Wojcik2015May}.
As a consequence, since the orbital degeneracy is 
not affected, one observes the orbital-$SU(2)$ Kondo effect along the lines
separating the charge states with occupation $(0,1)\leftrightarrow (1,0)$,
and $(1,2)\leftrightarrow (2,1)$ electrons,
as well as $(2,0)\leftrightarrow (1,1)$ and $(1,1)\leftrightarrow (0,2)$, see \fig{Fig:stability}(b).
Otherwise the conductance is generally suppressed
except for some special lines where $\Delta \varepsilon_{\rm exch}\approx 0$.

For a single quantum dot~\cite{Martinek2005Sep}, 
$\exch \approx (2p\Gamma / \pi) \log |\e / (\e+U)|$
and vanishes, i.e.  $\exch \approx 0$,
at the particle-hole symmetry point $\e = -U/2$.
In the absence of coupling between the two dots,
the Kondo effect in the first (second) dot
would be thus present for $\e_1 = -U/2$ ($\e_2 = -U/2$)
for any value of $\e_2$ ($\e_1$),
resulting in straight vertical and horizontal lines
in the $(\e_1,\e_2)$-plane
of the Kondo-enhanced conductance.
However, in the presence of capacitive coupling between the dots,
the lines become distorted by the inter-dot Coulomb correlations $U^\prime$,
as can be seen in \fig{Fig:stability}(b).

The difference in conductance in the two magnetic configurations
is reflected in the TMR, which is shown in \fig{Fig:stability}(c).
For transport regimes with even occupancy of each dot,
elastic cotunneling processes dominate the current and the TMR
is given by \cite{weymannPRB05} ${\rm TMR} \approx 2p^2/(1-p^2)$.
For odd occupancy, the Kondo effect is present in the case
of antiparallel configuration, while in the parallel configuration
it is suppressed by the exchange field, such that $G^{\rm P} \ll G^{\rm AP}$
and ${\rm TMR} \to -1$ \cite{weymannPRB11}.
On the other hand,
for such values of $\e_1$ and $\e_2$ that the exchange field 
vanishes, one has, $G^{\rm AP} / G^{\rm P} = 1-p^2$,
which yields ${\rm TMR} = p^2 / (1-p^2)$, a ratio which 
is valid irrespective of the  $SU(2)$ or $SU(4)$
Kondo regimes.

To understand the influence of ferromagnetic leads
on transport, in the following we will analyze the behavior
of the conductance and the TMR as function of spin polarization of the leads,
as well as temperature along different cuts in the stability diagram. 
We shall consider two such cross-sections defined as:
$(i)$ $\e_2 +\e_1 = - U'$ and $(ii)$  $\e_1 = \e_2$,
in the stability diagram.
In what follows we shall label them cut (line) 1 and 2. 

\subsection{Conductance and TMR along cross-sections}\label{sec:cut}

\begin{figure}[t]
  \includegraphics[width=0.85\columnwidth]{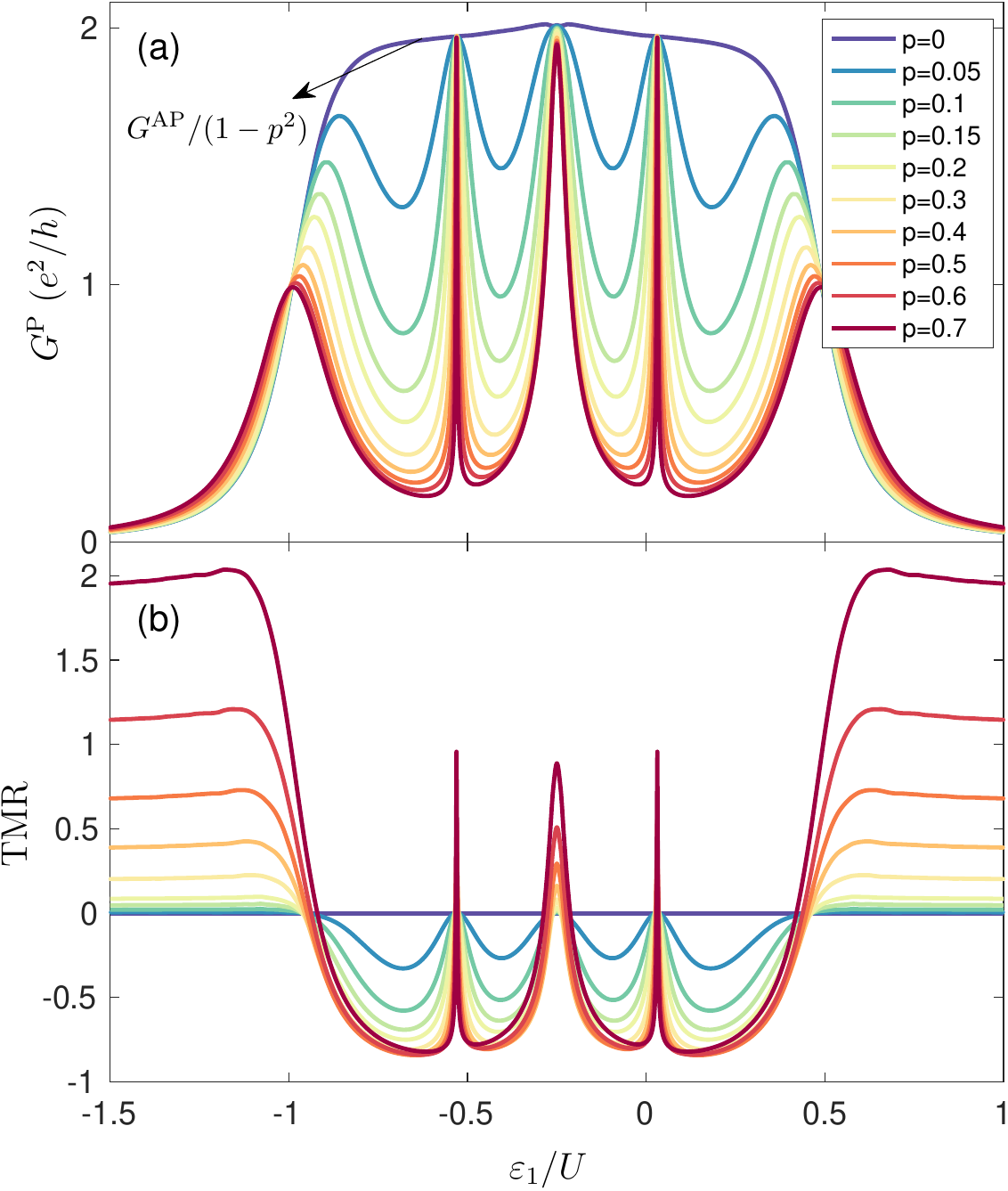}
  \caption{\label{Fig:cross_1}
  (a) The linear conductance in both magnetic configurations
  and (b) the resulting TMR as a function of $\e_1$ with $\e_2 = -\e_1 - U'$
  calculated for different values of leads' spin polarization $p$,
  as indicated.
  The conductance in the antiparallel configuration is given by
  the curve for $p=0$ multiplied with a factor of $(1-p^2)$, cf. \eq{Eq:GAP}.
  The parameters are the same as in \fig{Fig:stability}.}
\end{figure}

The linear conductance in both magnetic configurations and the TMR
calculated as a function of  $\e_1$ with $\e_2 +\e_1 = - U'$
for different values of spin polarization $p$ are shown in \fig{Fig:cross_1}.
By changing the level position, the occupation of the DQD
changes from $(2,0)$ for $\e_1\lesssim -U$, to $(1,0)$ for $-U\lesssim \e_1\lesssim -U'/2$,
$(0,1)$ for $-U'/2\lesssim\e_1\lesssim U/2$, and to ($0,2$) for $\e_1\gtrsim U/2$.
In the nonmagnetic lead case, in the odd occupancy regime
the regular spin-$SU(2)$ Kondo effect develops 
with conductance reaching $\approx 2e^2/h$, see \fig{Fig:cross_1}(a).
Moreover, for $\e_1 = -U'/2$, an additional orbital degeneracy 
occurs and the system exhibits the $SU(4)$ Kondo effect,
but the conductance remains $G\approx 2e^2/h$. These different types of the Kondo effects
are hardly distinguishable by the conductance itself when $T\ll \{T_K^{SU(2)}, T_K^{SU(4)}\}$,
as it remains close to the unitary value, $G\approx 2e^2/h$ in the whole singly occupied DQD regime,
see \fig{Fig:cross_1}(a).
However, they can be revealed at larger temperatures, i.e. $T\gtrsim \{T_K^{SU(2)}, T_K^{SU(4)}\}$
or in the case of ferromagnetic leads.

When $p>0$, the conductance gets modified.
The behavior in the AP configuration is still featureless, 
similar to the case of normal leads as $G^{\rm AP}(p) = G^{\rm AP}(p=0)(1-p^2)$. 
However, the conductance in the P configuration
reveals a nontrivial interplay between the spin-resolved DQD level renormalization
and the correlations bringing about the Kondo effect.
With increasing the spin polarization, the strength of the exchange field
increases and once $|\exch|$ becomes larger than the corresponding Kondo scale,
the conductance drops.
This can be observed in the whole odd occupation regime 
shown in \fig{Fig:cross_1}, i.e. for $-U\lesssim \e_1\lesssim U/2$, except for some special
values of the level position where, again, $\exch\approx 0$.
For $\e_1 \approx -U/2$, the exchange field in the first dot
vanishes, while for $\e_1 = 0$ (corresponding to $\e_2\approx-U/2$)
the exchange field in the second dot vanishes.
As a result, the total conductance reveals two peaks
for $\e_1 \approx \{ -U/2, 0 \}$
with an almost unitary conductance $G^{\rm P} \approx 2e^2/h$.
The height of these peaks remains almost constant, but their 
width depends on $p$, 
as the exchange field
increases with $p$,
and a smaller detuning is needed for the condition $|\exch|\gtrsim T_K^{SU(2)}$ to be fulfilled.
In addition, a spin-polarization independent resonance is also present 
for $\e_1 = -U'/2$ (note that then $\e_2 = \e_1$).
This is exactly the special point we have analyzed in Sec.~\ref{sec:SU4}
that shows the $SU(4)$ to $SU(2)$ crossover.
Although the maximum value of conductance does not depend at this point
on the polarization $p$, the system's ground state does change.
For $p=0$, it exhibits four-fold degeneracy,
which becomes reduced to two-fold degeneracy 
when increasing spin polarization.
Consequently, the $SU(4)$ Kondo effect becomes
reduced to the orbital $SU(2)$ Kondo effect
once $|\exch| \gtrsim T_K^{SU(4)}$.
The width of the resonance for $\e_1 \approx -U'/2$  is determined by the 
condition $|\Delta \e| \approx T_K^{SU(2)}$~\cite{fn7},
where $\Delta\e = \e_2 - \e_1$ corresponds to
the pseudo-Zeeman splitting.

\begin{figure}[!tb]
  \includegraphics[width=0.85\columnwidth]{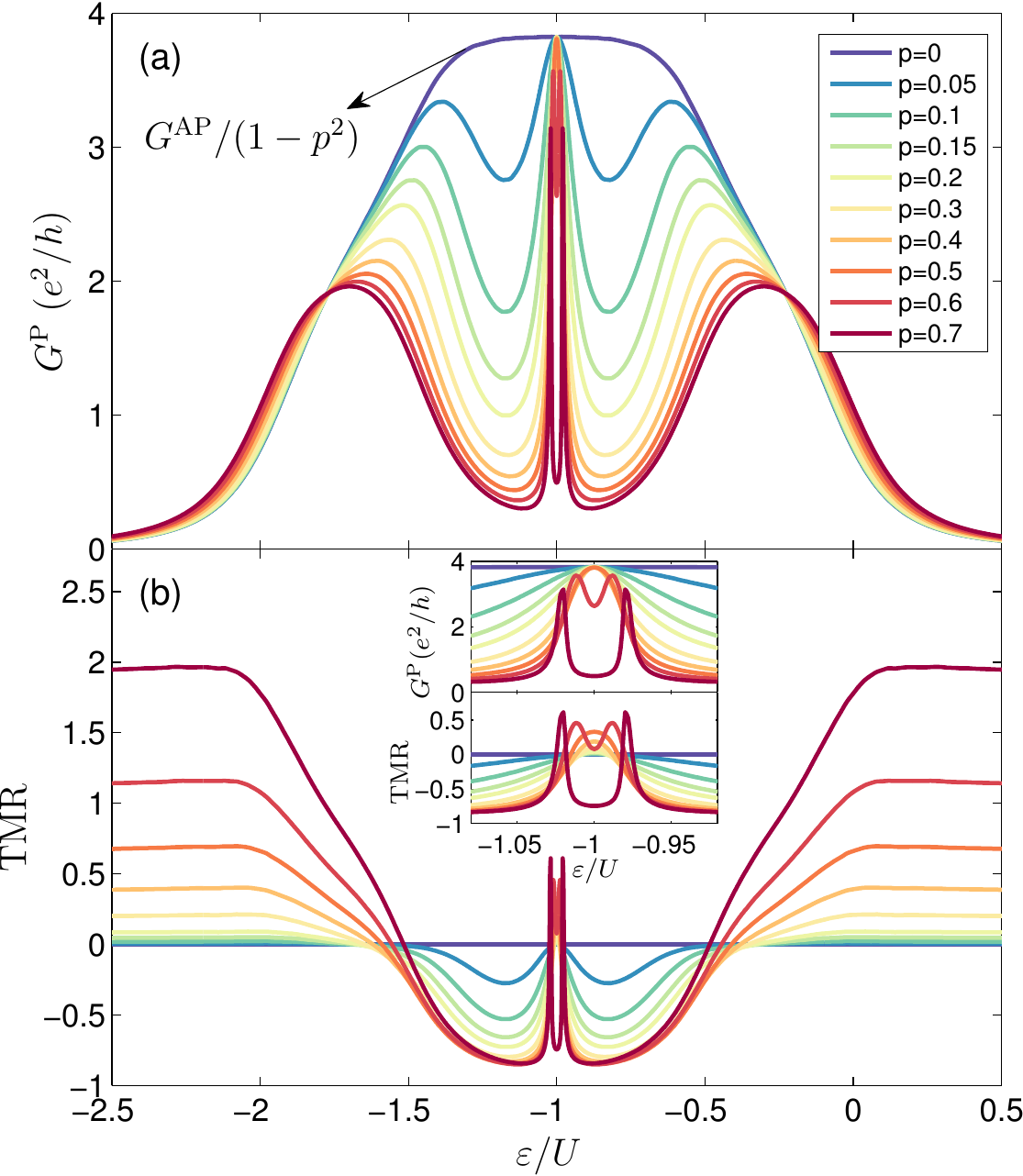}
  \caption{\label{Fig:cross_2}
  (a) The linear conductance in both magnetic configurations
  and (b) the resulting TMR as a function of $\e_1 = \e_2\equiv \e$
  calculated for different values of leads' spin polarization $p$, as indicated. 
  The inset shows the zoom into the transport regime
  around $\e =  -U/2-U'$.
  The conductance in the antiparallel configuration is given by
  the curve for $p=0$ multiplied with a factor of $(1-p^2)$, cf. \eq{Eq:GAP}.  
  The parameters are the same as in \fig{Fig:stability}.}
\end{figure}

The $\e_1$-dependence of the TMR for different spin polarizations 
along the first cut we consider 
is shown in \fig{Fig:cross_1}(b). The transport regimes
discussed above are clearly visible. In the even occupation regime
the TMR is given by ${\rm TMR} = 2p^2 / (1-p^2)$,
while in the case of odd DQD occupation, the TMR 
drops to ${\rm TMR} =-1$ with increasing $p$,
except for $\e_1 = -U/2$, $\e_1 = -U'/2$ and $\e_1 = 0$,
where ${\rm TMR} = p^2 / (1-p^2)$.

Let us now analyze the transport behavior along the second cut, 
where $\e_1 = \e_2 \equiv  \e$. Along this line, when $\e\gtrsim 0$,
the DQD is empty, for $-U'\lesssim \e\lesssim 0$ it is singly occupied,
for $-U-U'\lesssim \e\lesssim -U'$ two electrons occupy the DQD,
when $-2U\lesssim \e\lesssim -U-U'$ there are three electrons in the DQD,
while for $\e\lesssim -2U$ the DQD is fully occupied with four electrons.
In the odd occupation regime, the ground state
has four-fold degeneracy and the system exhibits
the $SU(4)$ Kondo effect in the case of nonmagnetic leads.
A plateau of $G\approx 2e^2/h$
associated with the $SU(4)$ Kondo effect is hardly visible
as a function of $\e$,
see the curve for $p=0$ in \fig{Fig:cross_2}(a).
This is because of a relatively large $\Gamma/U$ ratio
considered in calculations
and the usual spin $SU(2)$ Kondo effect, which develops
in both quantum dots yielding $G = 4e^2/h$ in the two-electron regime
in the case of $p=0$.
For finite $p$, in the parallel configuration the conductance becomes
however suppressed, except for $\e \approx -U/2-U'$, cf. \fig{Fig:stability}(b), where 
the exchange field cancels and the Kondo phenomenon can develop.
Moreover, the two plateaus in the odd-electron regime,
associated with the orbital $SU(2)$ Kondo effect, are clearly visible,
see e.g. the case of $p=0.7$ in \fig{Fig:cross_2}(a).
This confirms that for $p=0$, i.e. in the absence of 
level spin-splitting, the ground state of the system was indeed
four-fold degenerate.

Another feature in the $\e$-dependence of the conductance
can be seen around $\e = -U/2 - U'$ for finite $p$, see \fig{Fig:cross_2}(a).
As already mentioned, when $\e \approx -U/2 - U'$, the exchange field
vanishes and one should observe the Kondo effect.
However, instead of a peak at $\e \approx -U/2 - U'$,
with increasing $p$, a dip develops with 
two small satellite peaks.
This effect is associated with an interplay between finite temperature,
exchange field and the Kondo temperature.
First of all, one should note that exchange field
can be tuned not only by changing the DQD levels
(by inducing detuning from $\e = -U/2 - U'$),
but it also grows with spin polarization \cite{Martinek2005Sep}.
Thus, for larger $p$, a smaller detuning from the point
$\e = -U/2 - U'$ is needed to suppress the Kondo-enhanced conductance,
see the width of $G^P$ in the inset in \fig{Fig:cross_2}.
On the other hand, increasing the spin polarization results in lowering of the corresponding
Kondo temperature \cite{martinekPRL03}
and, once $T_K\lesssim T$, the conductance becomes suppressed at $\e = -U/2 - U'$.
The crucial observation is that $T_K$
also depends on detuning from the particle-hole symmetry point $\e = -U/2 - U'$
and grows with increasing this detuning.
As a consequence, small side peaks, on either side of 
$\e = -U/2 - U'$, develop in $G^P$ for
such values of $\e$ that $T_K \approx T$.
Note that these peaks are visible as long as $T_K \gtrsim |\exch|$,
and once this condition is not met any more,
which happens for even larger $p$, $G^P$ becomes suppressed.

The corresponding dependence of the TMR  is shown in \fig{Fig:cross_2}(b).
In this figure one can clearly identify all the TMR values
discussed earlier. In the empty and fully occupied DQD regime,
the elastic cotunneling gives rise to ${\rm TMR} = 2p^2/(1-p^2)$.
In the odd occupation regime, the TMR value
drops by a factor of $2$, while in the case of 
$-U-U'<\e<-U'$ the TMR is generally suppressed by the exchange field,
${\rm TMR} \to -1$, except for the middle of the Coulomb diamond, i.e. around $\e = -U/2 - U'$.
There, for large spin polarization,
the TMR displays two peaks on either side of $\e = -U/2 - U'$,
see the inset in \fig{Fig:cross_2}(b), resulting from the corresponding peaks in $G^P$.
The finite temperature effects visible in \fig{Fig:cross_2} lead our discussion
to the analysis of transport properties at different temperatures.
This is presented in the next section.

\subsection{Finite temperature effects}

\begin{figure}[t]
  \includegraphics[width=0.85\columnwidth]{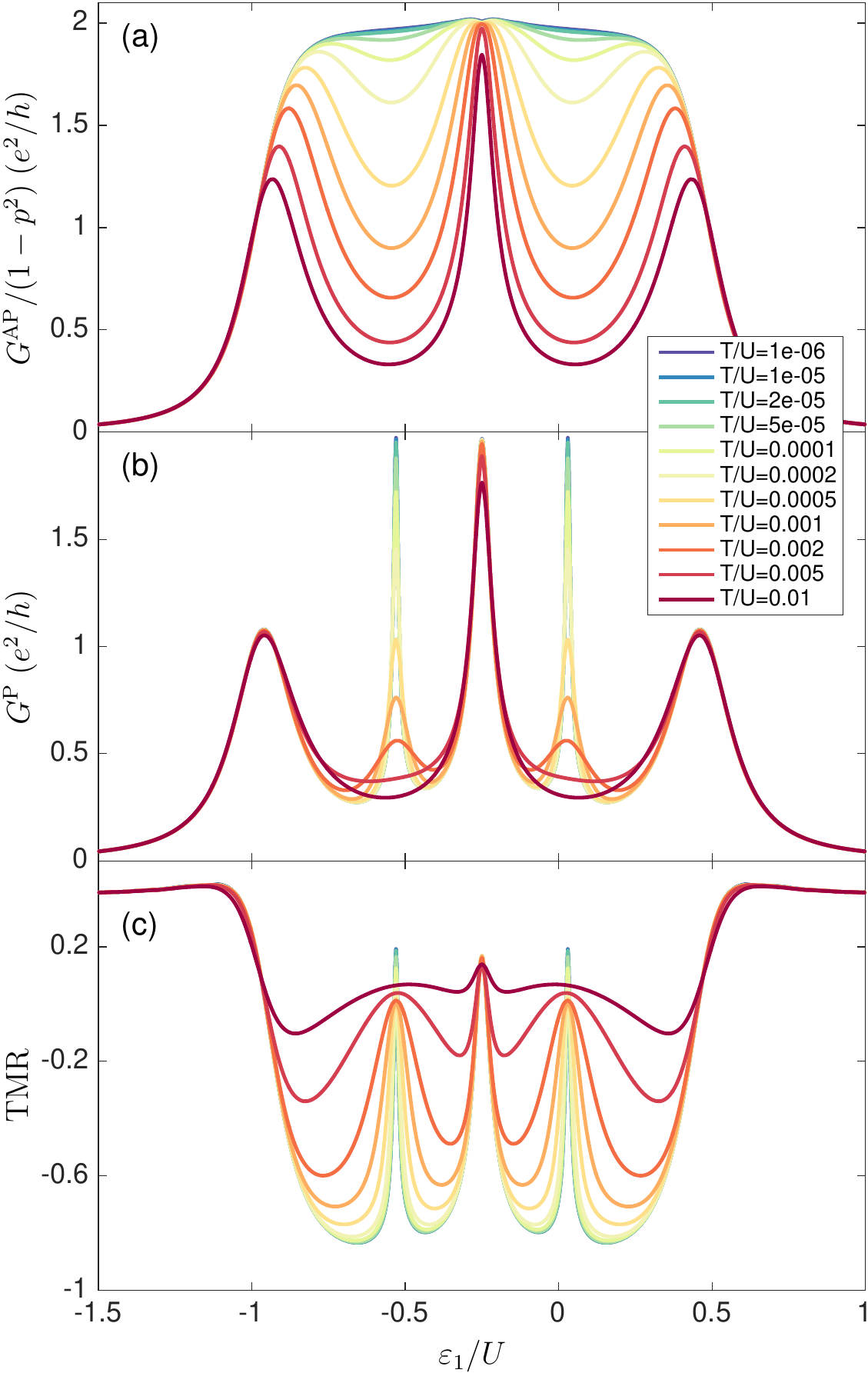}
  \caption{\label{Fig:cross_T_1}
  The linear conductance in (a) the antiparallel and (b) parallel magnetic configurations,
  as well as (c) the resulting TMR as a function of $\e_1$ with $\e_2 = -\e_1 - U'$
  calculated for different temperatures, as indicated.
  The other parameters are the same as in \fig{Fig:stability}.}
\end{figure}

\begin{figure}[t]
  \includegraphics[width=0.85\columnwidth]{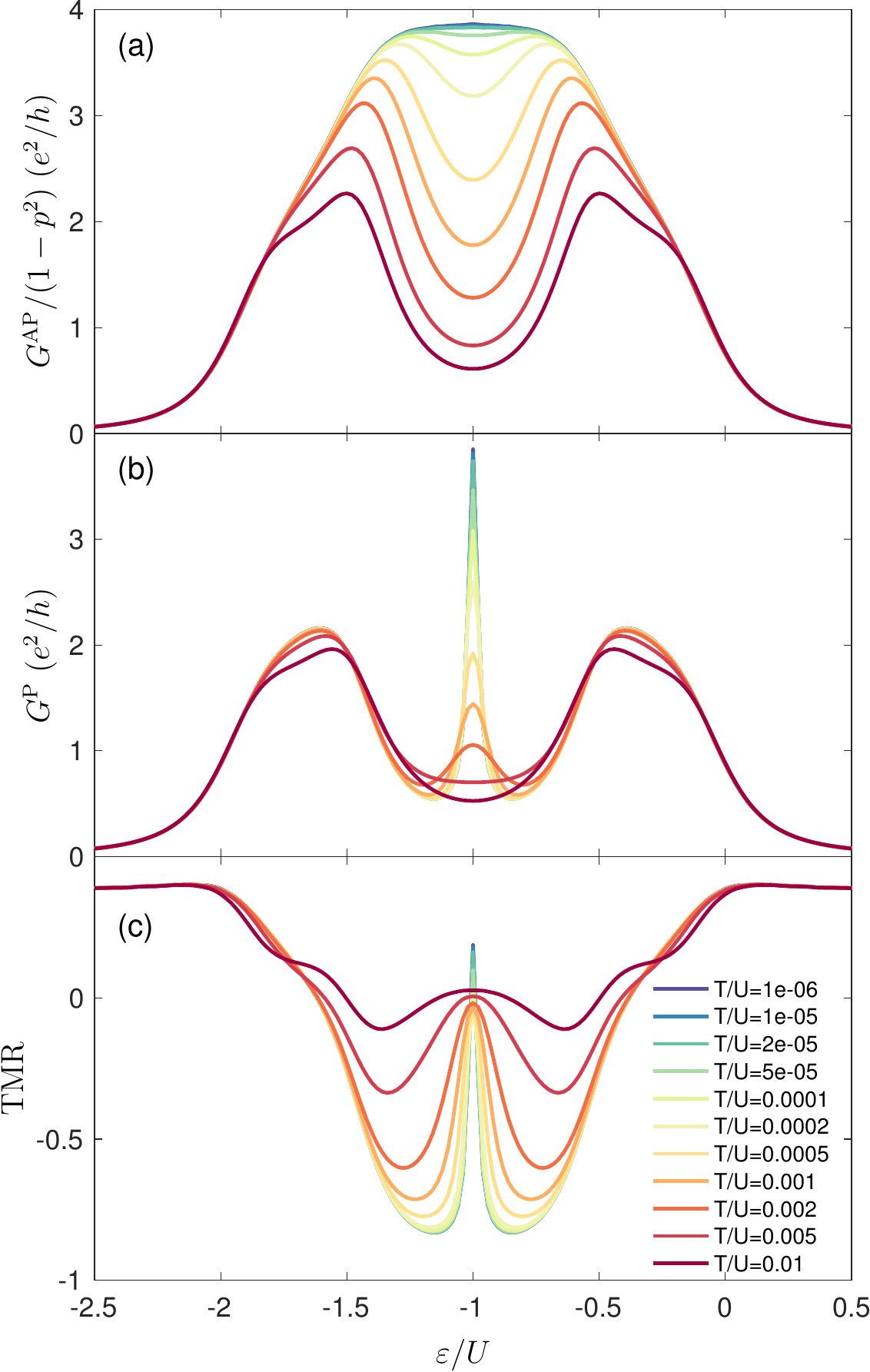}
  \caption{\label{Fig:cross_T_2}
  The linear conductance in (a) the antiparallel and (b) parallel magnetic configurations,
  as well as (c) the resulting TMR as a function of $\e_1 = \e_2\equiv \e$
  calculated for different temperatures, as indicated.
  The other parameters are the same as in \fig{Fig:stability}.}
\end{figure}

In this section we discuss the effect of the temperature on the linear 
conductance and TMR. For that we evaluated the conductance in both
AP and P magnetic configurations at various temperatures along the two cuts 
discussed in Sec.~\ref{sec:cut}. In Fig.~\ref{Fig:cross_T_1} we display 
the evolution of the conductance along the first cross-section,  $\e_1$ with $\e_2 +\e_1 = - U'$. 

At low temperatures, i.e. $T\lesssim\{ T^{SU(4)}_K , T^{SU(2)}_K\}$, 
the conductance in the antiparallel
configuration exhibits a plateau in the singly occupied DQD
transport regime~\cite{fn8}. This plateau changes when the temperature is increased.
First, the conductance becomes suppressed in 
the $SU(2)$ Kondo regime, and 
at some intermediate temperature,
$T^{SU(4)}_K\gtrsim T \gtrsim T^{SU(2)}_K$,
the resonances at $\e_1 \approx -U$
and $\e_1 \approx U/2$ survive,
together with the $SU(4)$ Kondo peak at $\e_1 \approx -U'/2$. 
From their temperature dependence
one can also estimate the Kondo temperatures:
In the middle of the spin $SU(2)$ Kondo valley and
for parameters assumed in \fig{Fig:cross_T_1}(a)
one finds, $T_K^{SU(2)}/U \approx 8.96 \cdot 10^{-4}$,
while the $SU(4)$ Kondo temperature for $\e_1 \approx -U'/2$ is, $T_K^{SU(4)}/U \approx 0.044$.

On the other hand, the evolution of $G^{\rm P}(T)$ along the first cut is 
completely different: The Kondo plateau is not present at low temperatures,
but only some narrow peaks occur at some specific values of $\e_1$.
It is obvious that the ones occurring at $\e_1\approx-U/2$ and $\e_1\approx 0$ 
are associated with the spin-$SU(2)$ Kondo effect~\cite{fn9}.
Note that in the case of finite $p$, the Kondo temperature
decreases with increasing spin polarization~\cite{martinekPRL03}.
Although, based on the previous analysis, 
we can safely attribute the feature at $\e_1 \approx -U'/2$
to the $SU(4)$ Kondo effect, from the evolution of $G^{\rm P}$ itself 
it is not that straightforward 
to decide what type of  correlations causes the conductance enhancement:
If $|\exch| \lesssim T_K^{SU(4)}$, then the $SU(4)$ nature of the ground state
is relevant, whereas for $|\exch| \gtrsim T_K^{SU(4)}$, the spin degeneracy is
 lifted and only the orbital
degrees of freedom are degenerate, resulting in orbital Kondo effect.
In fact, for parameters assumed 
in \fig{Fig:cross_T_1}(b), the strength of the exchange field
is comparable to $T_K^{SU(4)}$.

The effects of finite temperature on transport behavior
along the second cut we considered  ($\e_1 = \e_2 \equiv  \e$) 
are presented in \fig{Fig:cross_T_2}.
In the case of antiparallel configuration,
the conductance in the middle Coulomb blockade regime
becomes quickly suppressed with increasing temperature.
However, in the $SU(4)$ Kondo regime,
the dependence of $G$ on $T$ is weak
in the considered temperature range, since even for
the highest temperature considered
$T \lesssim T_K^{SU(4)}$.
A similar tendency can be observed 
in the case of parallel alignment.
A strong temperature dependence is only
revealed for the Kondo peak 
at $\e_1 = -U/2 -U'$,
while in the other transport regimes the linear conductance 
only weakly depends on $T$.

Finally, the TMR evaluated at various temperatures along the two cross-sections 
is shown in Figs. \ref{Fig:cross_T_1}(c) and \ref{Fig:cross_T_2}(c).
In these figures one can clearly identify
all the TMR values discussed earlier.
The general conclusion is that with increasing
the temperature, TMR extrema become suppressed,
such that in the very high temperature limit ($T \gtrsim U$, not shown),
the TMR would be  independent of 
$\e_1$ and $\e_2$, i.e. ${\rm TMR} \approx p^2/(1-p^2)$~\cite{weymannPRB11}.

\subsection{Ferromagnets with different coercive fields}

\begin{figure}[t!]
  \includegraphics[width=0.9\columnwidth]{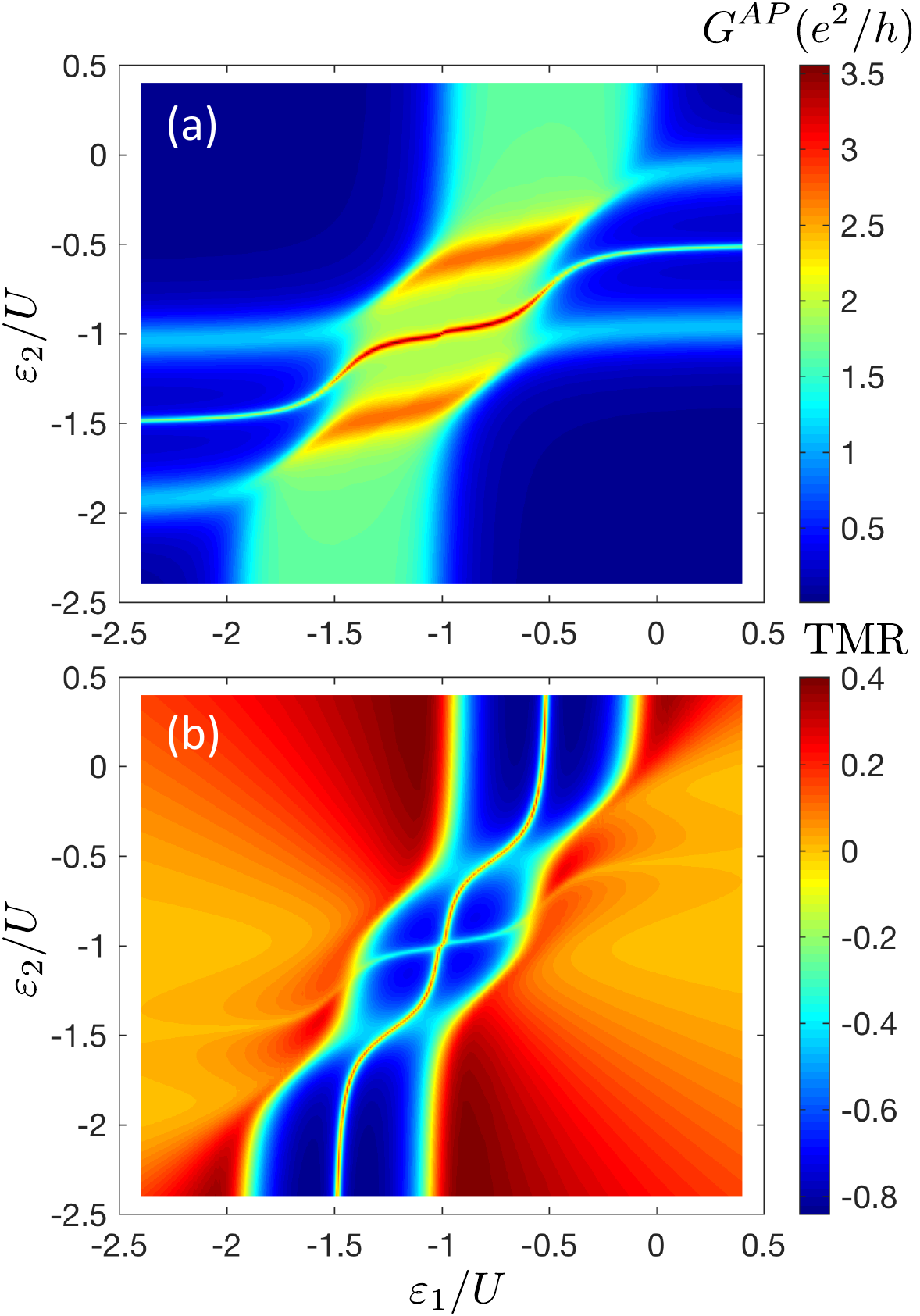}
  \caption{\label{Fig:mixed}
  (a) The linear conductance in the {\it mixed} antiparallel configuration
  and (b) the TMR as a function of DQD energy levels.
  The parameters are same as in \fig{Fig:stability}.
  In the {\it mixed} antiparallel configuration, the magnetization
  of one of the leads attached to the first dot 
  is opposite to the other leads' magnetizations.}
\end{figure}

In this section we discuss the magnetoresistive properties of the 
device assuming an experimentally relevant situation,
when the coercive fields of the ferromagnetic electrodes are different.
For sufficiently strong magnetic field (but still much smaller 
then the field necessary to induce a considerable Zeeman splitting),
the magnetizations of all electrodes are aligned (parallel configuration).
So far in our analysis
we have assumed that there is a difference between coercive
fields of the left and right electrodes, such that at certain field
the leads on one side of the junction flip their magnetizations and
the antiparallel configuration occurs, see \fig{fig:sketch}.
However, it may happen that only one of the electrodes flips its magnetic moment,
resulting in a {\it mixed} antiparallel configuration: For example
the leads coupled to the first dot are in the antiparallel,
while the leads attached to the second one are in the parallel magnetic configuration. 
The transport characteristics for such a situation are shown in \fig{Fig:mixed}.
One can still identify charged stability regions separated by lines 
with large conductance: When changing $\e_1$ the system exhibits
a Kondo plateau (visible in the transport regions for $\e_2\lesssim -U-U'$ and $\e_2\gtrsim -U')$,
while as a function of $\e_2$ the characteristic
suppression of the Kondo resonance by the exchange field occurs, see 
\fig{Fig:mixed}(a). The total conductance shows then an enhancement to $G^{\rm AP} = 4e^2(1-p^2/2)/h$
for such position of the DQD levels that the exchange field on the second dot vanishes.
The whole DQD level dependence of conductance in the {\it mixed}
configuration can be understood based on the analysis presented in Sec.~\ref{sec:cut},
and it results in the associated behavior of the TMR, which is shown in \fig{Fig:mixed}(b).

\section{Conclusions}
\label{sec:conclusions}

In this paper we studied the linear-response transport properties of
double quantum dot system coupled to ferromagnetic leads in the Kondo regime.
The emphasis was put on the transport regime where the system exhibits
the $SU(4)$ Kondo effect, which was thoroughly studied against
different material parameters of ferromagnetic contacts
and magnetic configurations of the device.
The calculations were performed with the non-perturbative
numerical renormalization group method and supplemented 
by an RG  analysis to describe the $SU(4)$ to $SU(2)$ crossover.
We demonstrated that the transport behavior becomes greatly modified
when the magnetic configuration of the device
changes from the antiparallel to the parallel one,
which is a direct consequence of the exchange field induced
DQD level splitting. This splitting generally breaks the 
spin-$SU(2)$ invariance, such that the system exhibits 
the orbital-$SU(2)$ Kondo effect in corresponding transport regimes.

We systematically investigated the evolution of the spectral functions
from the $SU(4)$ to the orbital-$SU(2)$ Kondo regime
upon increasing the leads' spin polarization $p$.
Interestingly, the corresponding Kondo temperature
reveals then a nonmonotonic dependence on $p$.
First, with increasing spin polarization, 
the Kondo temperature drops, which is related
to the reduction of the four-fold degeneracy to the two-fold one.
However, further increase of $p$ results 
in an enhancement of the orbital Kondo temperature,
such that for large spin polarization it may even exceed the $SU(4)$ Kondo temperature.
This behavior is completely different compared
to the single quantum dot case when monotonic dependence
of the spin-$SU(2)$ Kondo temperature on spin polarization
was predicted at the particle-hole symmetry point \cite{martinekPRL03}.

Finally, we also analyzed the magnetoresistive properties of the device
in the case when the ferromagnets have different coercive fields,
such that {\it mixed} antiparallel configuration is formed.
In such a case the transport behavior is a result
of contributions from the parallel and antiparallel configurations
of both quantum dots.

\section*{acknowledgements}
This work was supported by the National Science Centre
in Poland through the Project No. DEC-2013/10/E/ST3/00213 and
by the Romanian National Authority for Scientific Research and Innovation, UEFISCDI, 
project numbers PN-II-RU-TE-2014-4-0432 and PN-III-P4-ID-PCE-2016-0032.
Computing time at the Pozna\'n Supercomputing
and Networking Center is acknowledged.



\begin{thebibliography}{70}
\bibitem [{fn1()}]{fn1}%
  \BibitemOpen
  \href@noop {} {\bibinfo  {journal} {The $SU(4)$ symmetry can be constructed
  as $SU(4)=SU_{\rm spin}(2)\times SU_{\rm orbital}(2)$, as $H_{\rm DQD} $ is
  $SU(2)$ invariant in both the spin and orbital sectors}\ }\BibitemShut
  {NoStop}%
\bibitem [{fn2()}]{fn2}%
  \BibitemOpen
\bibfield  {journal} {  }\href@noop {} {\bibinfo  {journal} {For the system to
  remain $SU(4)$ invariant it is mandatory for the polarization of the leads to
  be zero and that there is no hopping between the dots}\ }\BibitemShut
  {NoStop}%
\bibitem [{fn3()}]{fn3}%
  \BibitemOpen
\bibfield  {journal} {  }\href@noop {} {\bibinfo  {journal} {In the present
  calculation we consider the fully symmetrical situation in which the
  tunneling amplitudes between each dot and the leads are all identical}\
  }\BibitemShut {NoStop}%
\bibitem [{fn4()}]{fn4}%
  \BibitemOpen
  \href@noop {} {\bibinfo  {journal} {At some particular points in the
  stability diagram the symmetry is higher in the spin and orbital space}\
  }\BibitemShut {NoStop}%
\bibitem [{fn5()}]{fn5}%
  \BibitemOpen
\bibfield  {journal} {  }\href@noop {} {\bibinfo  {journal} {To obtain the
  relevant spectral functions we use the usual log-Gaussian broadening kernel,
  \cite{WeichselbaumPRL07}, however the conductance is calculated directly from
  discrete data, \cite{WeymannPRB13}, which makes the results robust against
  broadening artifacts \cite{zitkoPRB09}}\ }\BibitemShut {NoStop}%
\bibitem [{fn6()}]{fn6}%
  \BibitemOpen
  \href@noop {} {\bibinfo  {journal} {Because the effect of finite $p$ in the
  antiparallel magnetic configuration is merely limited to a
  polarization-dependent prefactor, in this section we focus on the case of the
  parallel configuration only}\ }\BibitemShut {NoStop}%
\bibitem [{fn8()}]{fn8}%
  \BibitemOpen
\bibfield  {journal} {  }\href@noop {} {\bibinfo  {journal} {For the parameters
  that we used, $T^{SU(4)}_K < T^{SU(2)}_K$}\ }\BibitemShut {NoStop}%
\bibitem [{fn9()}]{fn9}%
  \BibitemOpen
\bibfield  {journal} {  }\href@noop {} {\bibinfo  {journal} {We can also use
  $G^{\rm P} (T)$ to estimate the Kondo scale. Here we estimate the Kondo
  temperature to be $T_K^{SU(2)}/U\approx 5.89\cdot 10^{-4}$}\ }\BibitemShut
  {NoStop}%
\end{thebibliography}


%

\end{document}